\documentclass[aps,pra,twocolumn,superscriptaddress,floatfix,showpacs]{revtex4}

\usepackage{graphicx}
\usepackage{color}
\usepackage{amsmath}

\begin{document}

\title{
Major role of multielectronic $K$-$L$ inter-shell resonant recombination processes\\
in Li- to O-like ions of Ar, Fe, and Kr
}

\author{C. Beilmann}
\affiliation{Max-Planck-Institut f\"ur Kernphysik, Saupfercheckweg 1, 69117 Heidelberg, Germany}
\author{Z. Harman}
\affiliation{Max-Planck-Institut f\"ur Kernphysik, Saupfercheckweg 1, 69117 Heidelberg, Germany}
\affiliation{ExtreMe Matter Institute (EMMI), Planckstra\ss e 1, 64291 Darmstadt, Germany}
\author{P. H. Mokler}
\affiliation{Max-Planck-Institut f\"ur Kernphysik, Saupfercheckweg 1, 69117 Heidelberg, Germany}
\author{S. Bernitt}
\affiliation{Max-Planck-Institut f\"ur Kernphysik, Saupfercheckweg 1, 69117 Heidelberg, Germany}
\author{C. H. Keitel}
\affiliation{Max-Planck-Institut f\"ur Kernphysik, Saupfercheckweg 1, 69117 Heidelberg, Germany}
\author{J. Ullrich}
\affiliation{Max-Planck-Institut f\"ur Kernphysik, Saupfercheckweg 1, 69117 Heidelberg, Germany}
\author{J. R. Crespo L\'opez-Urrutia}
\affiliation{Max-Planck-Institut f\"ur Kernphysik, Saupfercheckweg 1, 69117 Heidelberg, Germany}

\date{\today}

\begin{abstract}

Dielectronic and higher-order resonant electron recombination processes including a $K$-shell
excitation were systematically measured at high resolution in electron beam ion traps. Storing highly charged Ar, Fe, and Kr ions,
the dependence on atomic number $Z$ of the contribution of these processes to the total recombination cross section was studied
and compared with theoretical calculations. Large higher-order resonant recombination contributions are found, especially for systems
with $10<Z<36$. In some cases, they even surpass the strength of the dielectronic channel, which was hitherto presumed to be always the
dominant one. These findings have consequences for the modeling of high-temperatur plasmas.
Features attributed to inter-shell quadruelectronic recombination were also observed. The experimental data obtained for the He-like
to O-like isoelectronic sequences compare well with the results of advanced relativistic distorted-wave calculations employing
multiconfiguration Dirac-Fock bound state wave functions that include threefold and fourfold excitations.

\end{abstract}


\maketitle

\section{\label{intro} Introduction}

Electron-electron interaction is a fundamental aspect of both atomic and molecular structure and reactions.
It governs the level structure and the time dependence of atomic collision processes, and thus also the behavior
of laboratory and astrophysical plasmas. Understanding electronic correlations and their consequences is therefore
of great importance for the development of theory, and a good knowledge of those is crucial for all aspects of plasma diagnostics.

\begin{figure}[hbt]
\includegraphics[width = \columnwidth]{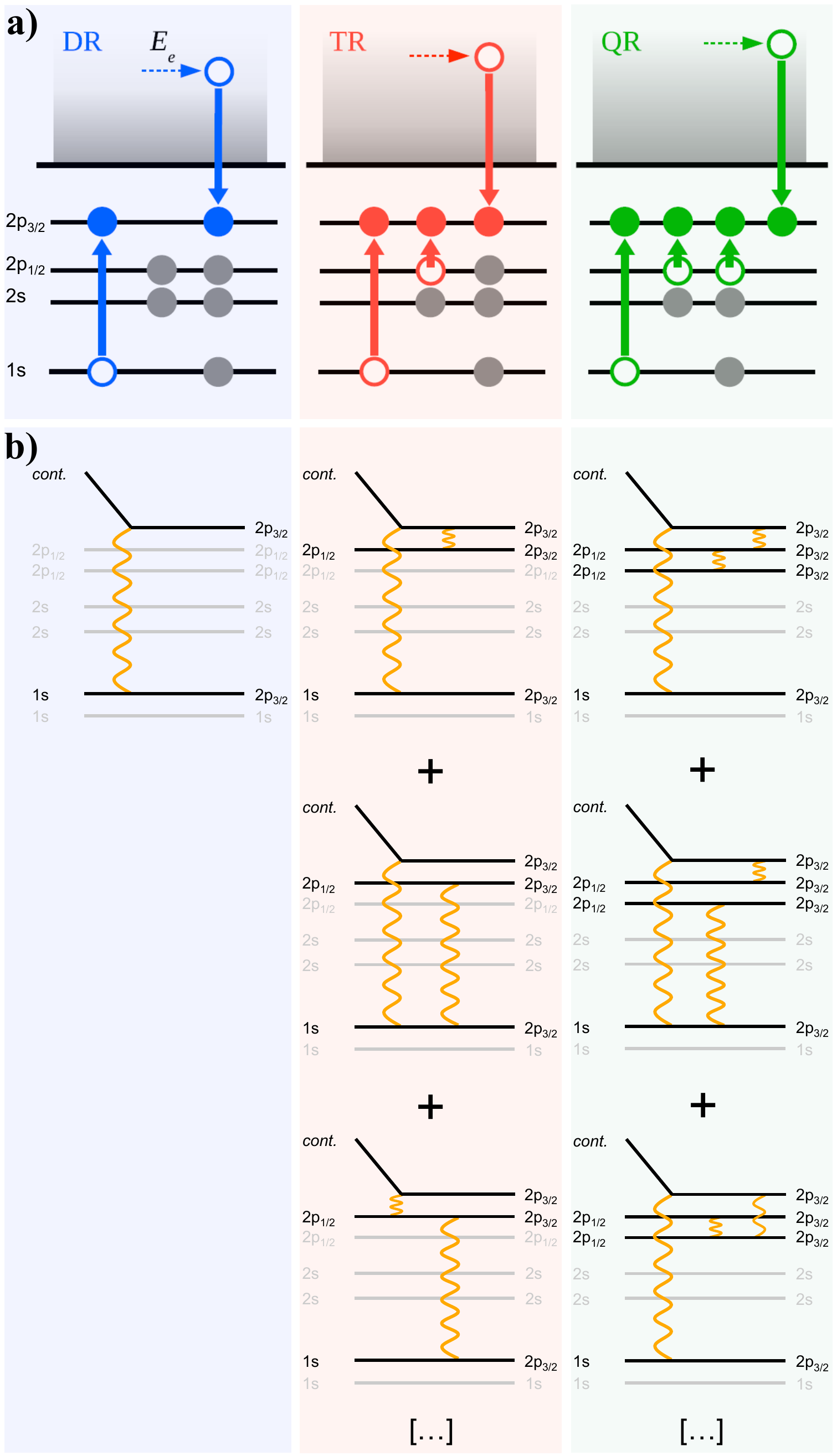}
\caption{\label{recombination-scheme}
(Color online) Schematic diagrams for first- and higher-order resonant electronic recombination (C-like ions):
dielectronic recombination (DR, left side in blue), trielectronic recombination (TR, middle panel in red),
and quadruelectronic recombination (QR, right side in green). At the top (a), energy level diagrams are shown.
In (b), the Feynman diagrams show some interaction possibilities.}
\end{figure}

Among the different processes involving electron-electron interaction, dielectronic recombination (DR) and its time reversal,
the Auger decay following photo-excitation of inner-shell electrons \cite{Massey42,Burgess64} are preeminent. In the first-order
DR process (see left side diagram in the top panel of Fig.~\ref{recombination-scheme}), a free electron is captured in
a collision with an ion, thereby transferring the sum of its energy to another bound electron and exciting it resonantly.
The ion is then left in a (usually) doubly-excited state which can decay radiatively, completing the DR process. Extended
experimental and theoretical investigations on DR with highly charged ions (HCI) have been carried out in recent years by
various groups (e.~g. \cite{Mueller08,Flambaum02,Gonzalez05,Harman06,Brandau08,Schippers09}).

Usually, DR is described in an independent-particle model, only taking into account the interaction between the two
electrons involved -- the initially free one and an active bound one. Higher-order electron correlations are not
accounted for in this approach. This is illustrated in the bottom diagram of Fig.~\ref{recombination-scheme} -- left side,
where the interaction process is represented by the Feynman diagram of one virtual photon exchange. Hence, the DR process
is usually denoted by the states of the electrons of concern; for instance, a $K$-$LL$ DR process (Auger notation,
cf.~\cite{Knapp95}) denotes the excitation of an inner-shell $K$ electron to the $L$-shell with a resonant capture
of the free electron also to the $L$-shell. However, our recent investigations have shown that higher-order correlations
with the active excitation of more than one electron may contribute considerably to recombination and affect plasma
parameters noticeably \cite{Beilmann09,Beilmann11a}. We focus on these higher-order effects in this paper.

Trielectronic recombination (TR) as an {\it intra-shell} process, where two bound $L$-shell electrons are excited within
their shell through the capture of one free electron, was reported as results of low-energy storage ring DR
measurements (cf.~\cite{Schnell03,Orban10,Schippers10}). In our previous experiments at electron beam ion traps (EBIT),
we found that the higher-order recombination process can also lead to far more energetic {\it inter-shell}
excitations~\cite{Beilmann09,Beilmann11a} with a rather large contribution. The corresponding process with $\Delta n=1$
excitation is exemplified in the middle diagram in Fig.~\ref{recombination-scheme} showing an inter-shell $KL$-$LLL$ TR
process, where a $K$- and an $L$-shell electron are excited simultaneously to vacant higher-lying $L$ levels under
resonant capture of the free electron, here also to the $L$-shell.

As more than two electrons are actively involved in TR, independent-particle models such as the (Dirac-)Hartree-Fock
method cannot adequately describe it. Electron correlations must be included using, e.~g., multiconfiguration or many-body perturbation theory methods.
Within the perturbative picture (at least) two virtual photons
must be exchanged for TR. Three out of six possible interactions are shown in the central part of Fig.~\ref{recombination-scheme}~(b).

Inter-shell TR in highly ionized Kr ions was observed and compared~\cite{Beilmann09} with extended calculations based
on the multiconfiguration Dirac-Fock (MCDF) procedure~\cite{Grant76,Harman06,Beilmann09}. In that work, the first indications
of third-order inter-shell quadruelectronic recombination (QR) were also found.
On the right side of Fig.~\ref{recombination-scheme}, inter-shell $KLL$-$LLLL$ QR is visualized. Here, electron capture to the $L$-shell
occurs simultaneously with the excitation of a $K$-shell electron, while two further $L$-shell electrons are also resonantly excited to
higher-lying $L$ states. Feynman diagrams with three exchanged virtual photons are shown in Fig.~\ref{recombination-scheme} (b)
(3 out of 24 possibilities are depicted.)

In our more recent work~\cite{Beilmann11a}, higher-order inter-shell recombination effects have also been reported for highly charged
Ar and Fe ions. We found that TR could overwhelm the first-order DR process for some ionic species in light elements, being the
dominant recombination channel in those systems.

In this paper we report in detail on our systematic high-resolution resonant recombination measurements performed at two of the
EBITs at the Max-Planck-Institut f\"ur Kernphysik in Heidelberg, Germany, namely, the HD-EBIT~\cite{Crespo99} and FLASH-EBIT~\cite{Epp07}.
We have extended the range of observed isoelectronic sequences, ranging now from He-like to O-like ions.

For the observation of the weak and blended higher-order structures, high-resolution experimental techniques are required.
Therefore, we give also insight into the forced evaporative ion cooling technique, essential for  high-resolution experiments.
With this method, lower-$Z$ elements, which require higher absolute resolution because of their lower fine-structure splittings,
become more readily accessible. This is mandatory for the aim of our investigations, because in the low-$Z$ regime, electron correlations are expected
to become stronger due to the relative decrease of the central Coulomb force compared to the $Z$-independent electron-electron interaction.

Our studies cover a range of elements relevant to different fields of
plasma physics: the study of Fe ions is central in astrophysics, and Ar as well as Kr ions have diagnostic uses e.g. in
fusion plasma devices~\cite{Widmann95,Bitter93}.

\section{\label{theo} Theoretical approach}

In a resonant recombination process, resonant electron capture is followed by the radiative
decay and stabilization of the ion's intermediate autoionizing state. It proceeds from an initial state $i$,
consisting of the ground-state ion and a continuum electron with an asymptotic momentum $\vec{p}$, through the intermediate autoionizing state $d$ to the bound final state $f$. 
The cross section for a specific resonant recombination channel is given (in atomic units) as a function of the electron's kinetic energy $E$ in terms of the resonance strengths $S_{i \to d \to f}$
as (see, e.~g.~\cite{Haan89,Zimmerer90,Zimmermann97})
\begin{eqnarray}
\label{eq:strength}
\sigma_{i \to d \to f}(E) &=& S_{i \to d \to f} L_d(E) \,,\\
S_{i \to d \to f} &=& \frac{2\pi^2}{p^2} \frac{2J_{d}+1}{2(2J_i+1)} \frac{\Gamma^r_{d \to f}
}{\Gamma_d} \Gamma^a_{d \to i} \,. \nonumber
\end{eqnarray}
Here, the Lorentzian line shape function
\begin{equation}
L_d(E) =  \frac{\Gamma_d/(2\pi)}
{(E-E_{\rm res})^2 + {\Gamma_d^2}/{4}}
\end{equation}
is normalized to unity on the energy scale, $\int{L(E)dE=1}$, and is centered around the resonance energy $E_{\rm res}$,
calculated as the difference of the intermediate and initial ionic level energies: $E_{\rm res}=E_d-E_i$.
$J_d$ and $J_i$ are the total angular momenta of the intermediate and the initial states of the recombination
process, respectively. The free electron momentum associated with the kinetic energy of
the initially free electron at resonance is given by $p=|\vec{p}|=\sqrt{(E_{\rm res}/c)^2 - c^2}$.
The total radiative width of the autoionizing intermediate state is $\Gamma^r_{d}=\sum_f \Gamma^r_{d \to f}$,
and $\Gamma^a_{d}=\sum_i \Gamma^a_{d \to i}$ is the total autoionization
width of the same state, summed over all possible Auger final states. $\Gamma_d$ denotes the total
natural line width of the resonant state: $\Gamma_d = \Gamma_d^r + \Gamma_d^a$.
The resonant capture rate is related to the rate of its time-reversed (Auger) process by the principle of
detailed balance and is defined, according to Fermi's golden rule, perturbatively as:
\begin{widetext}
\begin{eqnarray}
\label{eq:dr-rate}
\Gamma^{\rm capt}_{i \to d} = \frac{2J_{d}+1}{2(2J_i+1)} \Gamma^a_{d \to i} &=& \frac{2\pi}{2(2J_i+1)} \sum_{M_{d}} \sum_{M_i m_s}
\int \sin(\theta) d\theta d\varphi |\langle\Psi_d; J_d M_d \Pi_d  | V | \Psi_i E; J_i M_i \Pi_i, \vec{p} m_s\rangle|^2 \nonumber \\ 
&=& 2\pi \sum_{\kappa} |\langle\Psi_d; J_d \Pi_d || V || \Psi_i E; J_i \Pi_i j\pi; J_d \Pi_d\rangle|^2 \,, 
\end{eqnarray}
\end{widetext}
In this equation, the transition amplitude, i.~e., the matrix element of the sum of the Coulomb and Breit interactions
$V=V^C + V^B$, is calculated with the initial bound-free antisymmetrized product state
$| \Psi_i E; J_i M_i \Pi_i, \vec{p} m_s\rangle$ and the resonant intermediate state. After averaging over the
initial magnetic quantum numbers $M_i$, $m_s$ and the direction $(\theta,\phi)$ of the incoming continuum
electron, and after performing a summation over the magnetic quantum numbers $M_d$ of the autoionizing
state, one obtains the partial wave expansion of the reduced matrix elements, as given in the last line
of the equation (\ref{eq:dr-rate}). $\kappa$ is the relativistic angular momentum quantum number appearing in the decomposition
of the continuum scattering state~\cite{Eichler95}. The last line of Eq.~(\ref{eq:dr-rate}) also shows that, owing to the usual
Coulomb interaction (or, generally, electron-electron interaction) selection rules, the total angular momentum of the initial state
(consisting of the bound ionic electrons and the continuum electron partial wave) and that of the autoionizing state must be equal.
The same holds for the parities $\Pi$ (given as the product of the single-electron parities) of the initial and autoionizing states.

In the case of dielectronic recombination, the autoionizing state may be approximately described by a single
configuration state function. 
For example, for C-like ions initially in their ground sate, $|\Psi_i \rangle=|1s^2 2s^2 2p_{1/2}^2\rangle$, the $K$-$LL$ DR autoionizing states may be $|1s 2s^2 2p_{1/2}^2 2p_{3/2}^2, J_d \Pi_d\rangle$, with $J_d=\{1/2,3/2,5/2\}$ and $\Pi_d=+1$.
This notation expresses that a configuration state function is constructed from antisymmetrized single-particle
wave functions coupled to a given $J_d$. 
Such states and the corresponding energy levels may be calculated by the multiconfiguration Dirac-Hartree-Fock (MCDF) approach with additional quantum electrodynamic and mass shift corrections~\cite{Harman06,Gonzalez06,Parpia96}. 
The bound state wave functions obtained numerically may be used to construct continuum orbitals and the matrix elements entering Eq.~(\ref{eq:dr-rate})~\cite{Zimmerer90,Harman06,Zimmermann97,Zakowicz03}.

The transition amplitude of the three-body interaction accounting for trielectronic recombination into a state
$|1s 2s^2 2p_{1/2} 2p_{3/2}^3, J_d \Pi_d \rangle$ may be described in a perturbative approximation as
\begin{eqnarray}
\label{eq:tr-pert}
&&\frac{\langle 1s 2s^2 2p_{1/2} 2p_{3/2}^3, J_d \Pi_d  |V|1s 2s^2 2p_{1/2}^2 2p_{3/2}^2, J_d \Pi_d\rangle}{E_{\rm 'TR'}-E_{\rm 'DR'}}\\
&&\times\langle 1s 2s^2 2p_{1/2}^2 2p_{3/2}^2, J_d \Pi_d|V|1s^2 2s^2 2p_{1/2}^2 E;J_i j;J_d \Pi_d\rangle \nonumber
\end{eqnarray}
Here, $E_{\rm 'TR'}$ and $E_{\rm 'DR'}$ stand for the energy eigenvalues associated with the states
$|1s 2s^2 2p_{1/2} 2p_{3/2}^3, J_d \Pi_d\rangle$ and $|1s 2s^2 2p_{1/2}^2 2p_{3/2}^2, J_d \Pi_d\rangle$, respectively.
Note that since the operator $V$ is a two-body operator, i.e. it only changes the occupation numbers of two single-electron states,
the trielectronic process cannot be described by first-order perturbation theory,
\begin{eqnarray}
&& \langle 1s 2s^2 2p_{1/2} 2p_{3/2}^3, J_d \Pi_d  |V|1s^2 2s^2 2p_{1/2}^2 E;J_i \Pi_{i} j \pi;J_d \Pi_d\rangle \nonumber \\
&& =0\,.
\end{eqnarray}

Equation~(\ref{eq:tr-pert}) also expresses the symmetry requirement that a TR autoionizing state can only be populated with noticeable probability
if there is a DR state of the same total angular momentum and parity (i.e. the matrix element in the numerator is non-zero) and it is close in energy
(i.e. the denominator is not too large as compared to the numerator). This acts toward a suppression of $KL$-$LLL$-TR for initially Be-like ions~\cite{Beilmann09}:
starting with the Be-like initial state, characterized by the dominant configuration $1s^2 2s^2$, $KL$-$LLL$ TR may only populate
two-hole autoionizing states such as $1s2s2p^3$, which are of negative parity. The nearby $K$-$LL$ DR single-hole autoionizing states
are of the type $1s2s^22p^2$ and of positive parity; therefore, these DR and TR autoionizing states do not mix (i.~e., the ratio in the
first line of Eq.~(\ref{eq:tr-pert}) is zero because of the Coulomb interaction selection rules). Higher-lying DR states, i.~e.,
those of type $1s2s^22p3s$ may certainly possess negative parity and mix with the TR state, however, as such states are energetically
far away from the TR state, the mixing is low (i.e. the energy denominator in the first line of Eq.~\ref{eq:tr-pert} is large).

The TR cross section may nevertheless be sizable due to the mixing of the initial (ground) state of the ion with excited-state configurations:
since in the description of the, e.g., Be-like initial state configurations of the type $1s^2 2p^2$ are present, the transition from such a state to a
TR configuration $1s 2s 2p^3$ is possible by a single Coulomb interaction, giving rise to significant TR peaks~\cite{Beilmann12}. This demonstrates that for an adequate description
of resonant recombination processes, mixing effects in both the initial and intermediate states has to be taken into account.

The ratio of TR to DR amplitudes is given by the fraction in the first line of Eq.~(\ref{eq:tr-pert}).
At low $Z$, this fraction may approach unity since the electron interaction operator $V$ acts on the strongly overlapping
TR and DR state functions, and the interaction energy in the numerator lies close to the $(E_{\rm 'TR'} - E_{\rm 'DR'})$ splitting.
The TR capture rates may thus become comparable to the DR capture rates. Because of this, 
it is necessary to  treat the
trielectronic process non-perturbatively: the autoionizing state formed by the trielectronic process
is described as the linear combination
\begin{eqnarray}
\label{eq6}
| {\rm TR} \rangle &=& c_1 | 1s 2s^2 2p_{1/2} 2p_{3/2}^3, J_d \Pi_d \rangle \nonumber \\
&+& c_2 |1s 2s^2 2p_{1/2}^2 2p_{3/2}^2, J_d \Pi_d\rangle \,,
\end{eqnarray}
and the autoionizing state populated by dielectronic capture is represented as
\begin{eqnarray}
\label{eq7}
| {\rm DR} \rangle &=& \tilde{c}_1 | 1s 2s^2 2p_{1/2}^2 2p_{3/2}^2, J_d \Pi_d \rangle \nonumber \\
&+& \tilde{c}_2 |1s 2s^2 2p_{1/2} 2p_{3/2}^3, J_d \Pi_d\rangle \,.
\end{eqnarray}
The expansion coefficients may be obtained by diagonalizing the total Hamiltonian of the system in a
configuration interaction approach or by the MCDF method.

Note that in equations (\ref{eq6}) and (\ref{eq7}), the first configuration is not necessarily the dominant one;
this holds only true in the high-$Z$ limit where the $jj$-coupling scheme is applicable.
The question may arise whether it is meaningful to define the number of electrons excited in a correlated many-electron shell,
which may be described by strongly mixing configurations in the low-$Z$ limit.
We find that, since a state such as (\ref{eq6}) contains two electronic holes, one in the $K$ shell and one in the $L$ shell,
which may both decay separately by photon emission, the process leading to its population can indeed be termed as trielectronic recombination.

Equations (\ref{eq6}) and (\ref{eq7}) may be easily extended to the case when 3 or more levels sharing the same $J_{d}\Pi_{d}$ quantum numbers are energetically close.
Correlation effects may be accounted for more accuracy
by including further configurations of the same symmetry $J_d \Pi_d$ in a multiconfiguration approach.

\section{\label{experiment} Experimental procedure}

\begin{figure}[tb]
\includegraphics[width = \columnwidth]{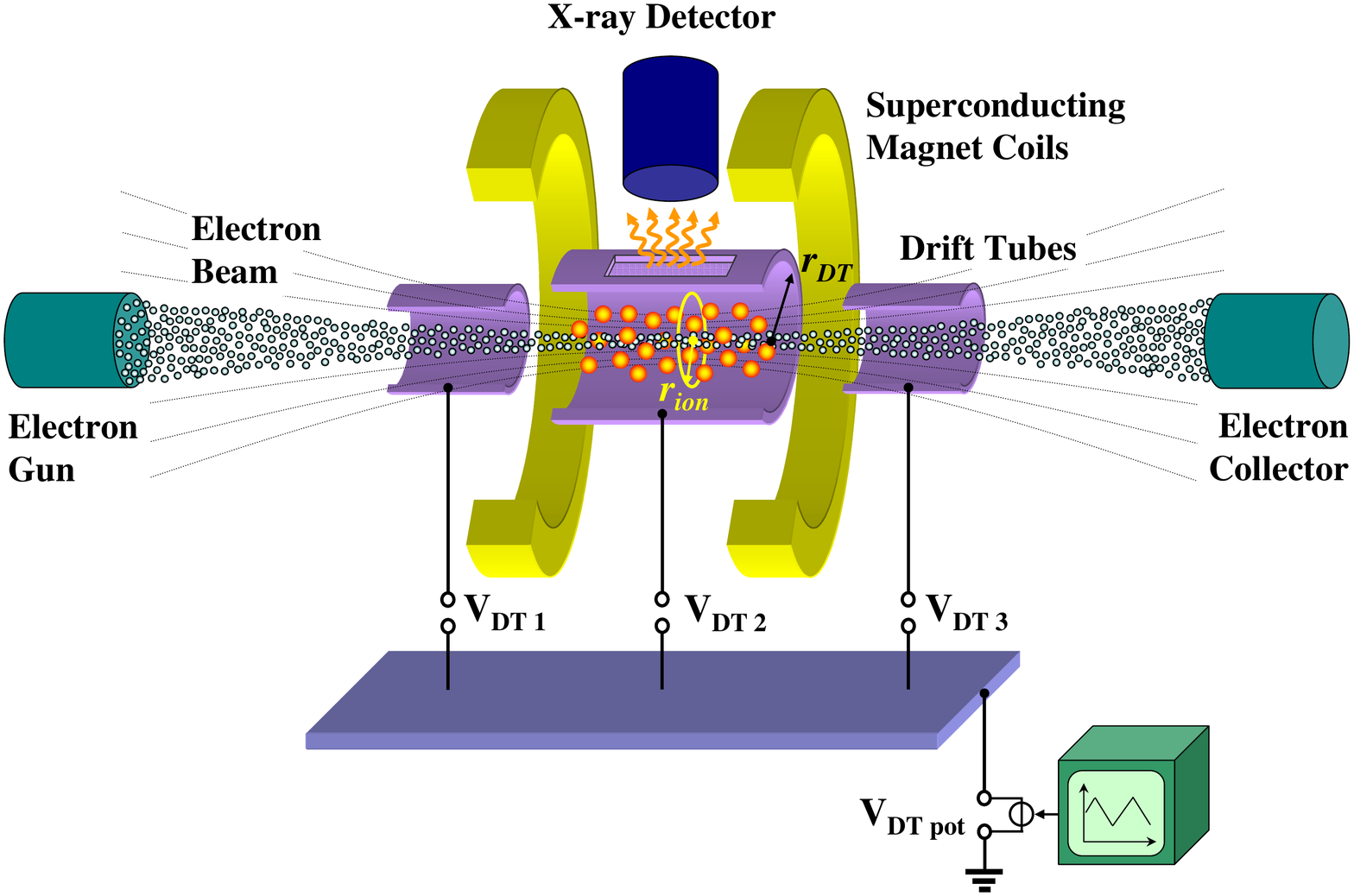}
\caption{\label{EBIT} (Color online) Principle of electron recombination experiments using an electron beam ion trap.
The whole drift tube assembly (including the central tube, V$_{\rm DT2}$) can be ramped by the common platform V$_{\rm DT pot}$.
Photons resulting from radiative stabilization are detected with an energy-dispersive germanium detector.}
\end{figure}

Resonant recombination was experimentally investigated by using electrons as projectiles and HCI 
as targets. The target ions are produced and confined in an EBIT by electron impact ionization of neutral target atoms. 
Electron beams with currents of 50 to 200 mA compressed to typical diameters of around 50 $\mu$m by a magnetic field of 6 to 8 T ensured a high current density in the reaction volume and thus a high ionization rate. 
This current density leads to a negative space charge potential that confines the positively charged ions in the radial direction around the electron beam axis. 
An arrangement of drift tubes allows for generating a potential well to trap the ions in the axial direction. 
This basic principle of an EBIT is displayed in Fig.~\ref{EBIT}. 

Projectile electrons are provided by the monoenergetic electron beam of the EBIT. 
Its energy is controlled by the voltage difference between the cathode of the electron gun $(V_{\text{cath}})$ and the central drift tube surrounding the reaction volume $(V_{\text{DT 2}})$. The variable bias acceleration voltage $V_{\text{DT pot}}$ is applied to all drift tubes, while keeping the ions trapped within a potential well of constant depth determined by the potential difference between the central drift tube and its immediate neighboring electrodes.
We have used the FLASH-EBIT \cite{Epp07} for the Ar and Fe experiments and the Heidelberg EBIT \cite{Crespo99} for the Kr measurements. 
The experimental procedure was the same in both devices.

For the experiments, we follow the widely applied scheme described in Ref.~\cite{Knapp93}. Here, we slowly vary the electron beam energy across a range of interest in
a sawtooth ramp with a slow slew rate of 2\,V/s in order to ensure steady-state conditions. Recombination involving $K$-$L$-excitation was monitored by measuring the
characteristic $K$ x-ray photons that are emitted by radiative stabilization of the $K$-hole excited state formed by resonant electron capture. At specific electron
energies, an enhancement of $K$-photon emission indicates a photorecombination resonance. A high-purity germanium x-ray detector is pointed at the reaction volume
side-on with respect to the electron beam through a 250-$\mu$m thick beryllium window. For the photon energy range of interest (from 2 to 14\,keV) this window is
nearly transparent. 

\begin{figure}[tb]
\includegraphics[width = \columnwidth]{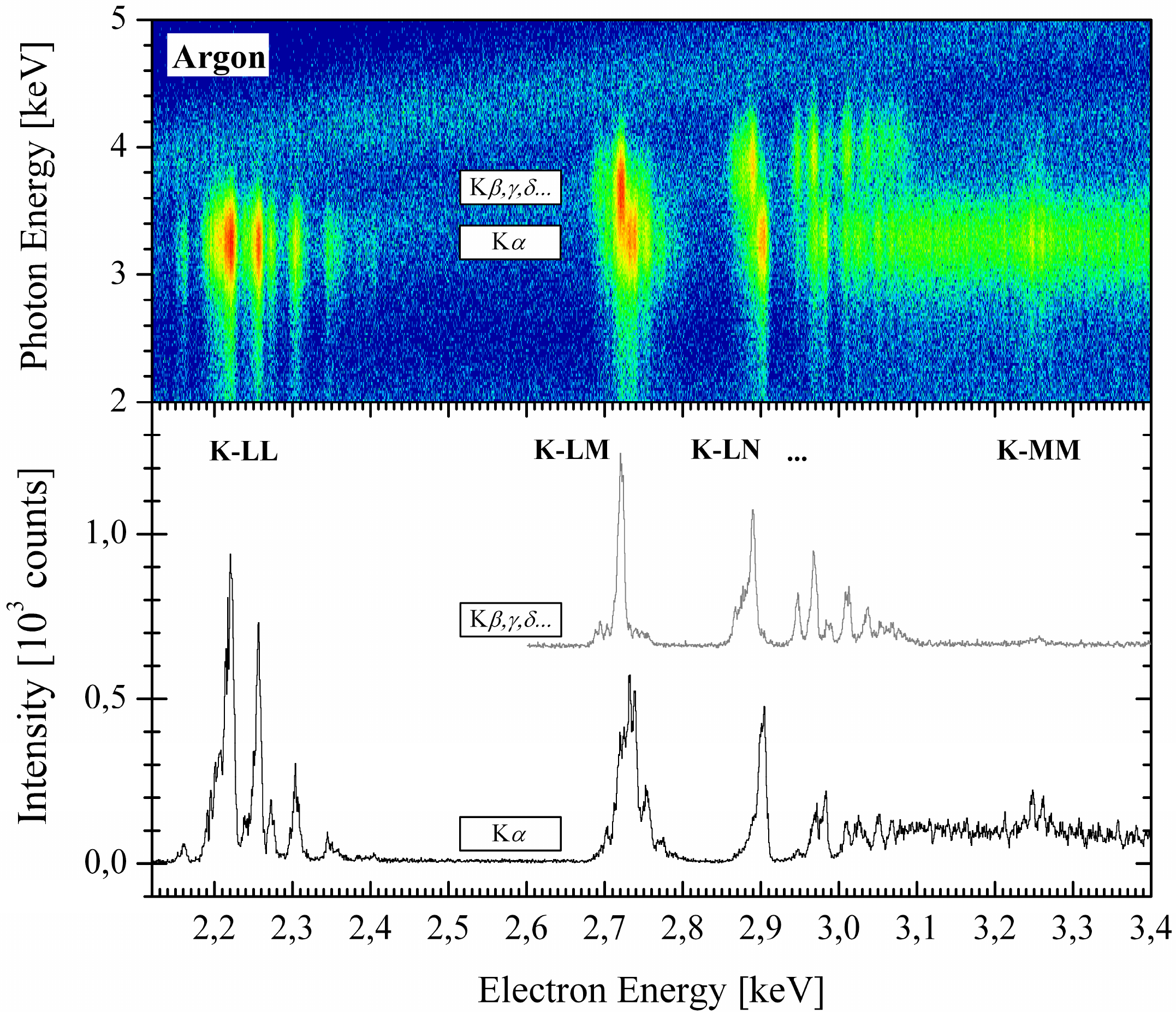}
\caption{\label{ar-overview} (Color online) Top (a): Photon intensity map
(vertical: photon energy; horizontal: electron energy) showing an overview of the photorecombination process for trapped Ar ions with open $L$-shell. The bright features correspond to resonances. Bottom (b): Projections onto the electron energy axis of stripes containing the $K_{\alpha}$ and $K_{\beta}$ emission lines (lower and upper spectrum, respectively).}
\end{figure}

A two-dimensional intensity plot in dependence of the electron beam energy and the emitted x-ray photon energy is displayed in
Fig.~\ref{ar-overview}\,a (top panel). The electron recombination resonances appear as bright spots. In this figure,
an overview of resonant recombination processes with $K$-shell excitation in He- to O-like Ar is presented, showing
both the $K$-$LL$ region this paper is focused on, and resonances for $K$-$LM$, $K$-$LN$, $\dotsc$, etc. recombination processes.
These resonances involving the $K$-shell and a higher-shell can also be seen as bright spots in the $K_\beta$ photon energy region.
In addition, $K$-$MM$ resonances can be identified at electron beam energies around 3.25\,keV. The projections shown below in
Fig.~\ref{ar-overview}\,b (bottom panel) represent the $K_{\alpha}$ emission ($K_{\alpha}$ cut) and $K_{\beta}$ and higher
emission as a function of the electron energy.

For the investigation of weak resonances, a good resolution on the electron energy is a prerequisite for separating them from stronger lines. 
In an EBIT, the resolution in recombination measurements is limited by the energetically broadened electron beam~\cite{Penetrante91,Penetrante91sc}. Earlier experiments used low electron beam currents for minimizing the absolute space charge potential, its width, as well as the ion heating caused by electron impact~\cite{Penetrante91sc}. 
However, a low beam current leads to a strong reduction of the ion production rate and also to a diminished recombination rate.

In our experiments we improved the resolution by forced evaporative cooling in combination with electron beam currents sufficient for an efficient ionization and abundant recombination yield. 
This was realized by lowering the axial potential applied to the drift tubes. To compensate a remaining trap generated by the space charge potential of the drift tubes having different diameters \cite{Beilmann10,Penetrante91sc}, a positive offset voltage has to be applied to the central drift tube referring to the surrounding ones. 
Under these settings, only a cold ion ensemble remains in the trap and the measurements can be performed with relatively high electron beam currents.

\begin{figure}[tb]
\includegraphics[width = \columnwidth]{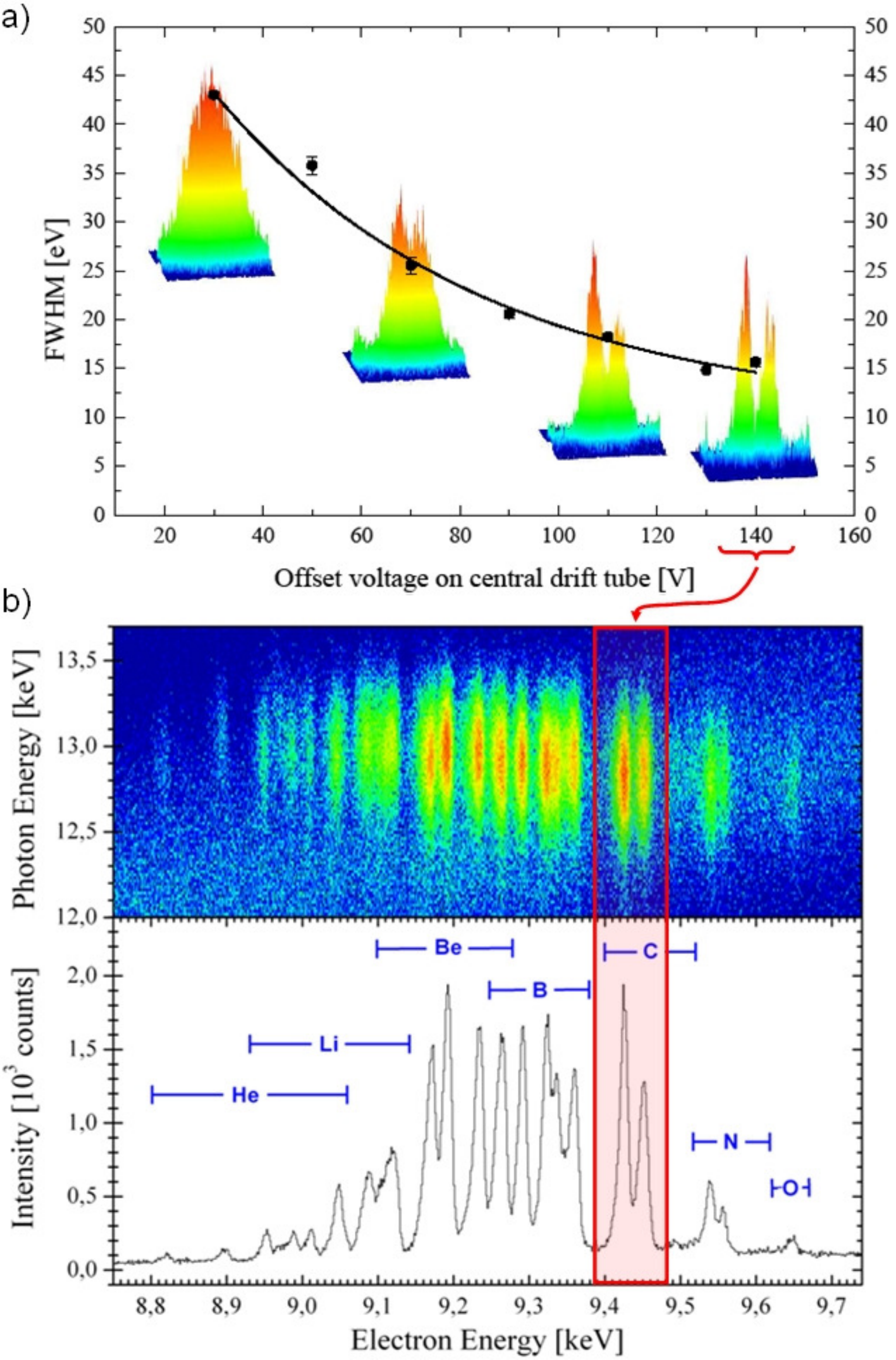}
\caption{\label{resolution-improvement} (Color online)
(a) Resolution improvement by evaporative cooling: Measured resolution (full width at half maximum, FWHM) for DR resonances in C-like Kr (around 9.44 keV)
plotted as function of the offset voltage  $\Delta V_{\rm DT2}$ of the central drift tube. The inserted diagrams give the
three dimensional intensity plots at offset voltages of 30, 60, 110 and 140 V. (b) Resulting photorecombination spectrum for
He- to O-like Kr ions. The region of interest for C-like ions displayed in part (a) of the figure is indicated; the regions for the different ionic sequences are labeled.}
\end{figure}

With this method, the resolution, defined as the ratio of the FWHM width to the space charge, was improved by roughly a factor of 5 in comparison with the
common measurement scheme~\cite{Beilmann10}.
Fig.~\ref{resolution-improvement}\,a demonstrates this advantage for two strong adjacent DR resonances in C-like Kr ions, namely, the
$([1s2s^{2} 2p_{1/2}^{2} (2p_{3/2}^{2})_{2}]_{5/2, (3/2 + 1/2)})$ lines. The resolution (full width at half maximum, FWHM) of the resonance with $J=5/2$
is plotted versus the voltage of the central drift tube with respect to the outer ones at an electron beam current of 200\,mA. The former unresolved
lines become clearly separated at the optimized point of cooling. In Fig.~\ref{resolution-improvement}\,b, a very well resolved recombination spectrum
allows for both a clear separation of charge state-specific resonances and the observation of otherwise blended weak features.

\section{\label{results} Results and discussion}

\begin{figure}[tb]
\includegraphics[width = \columnwidth]{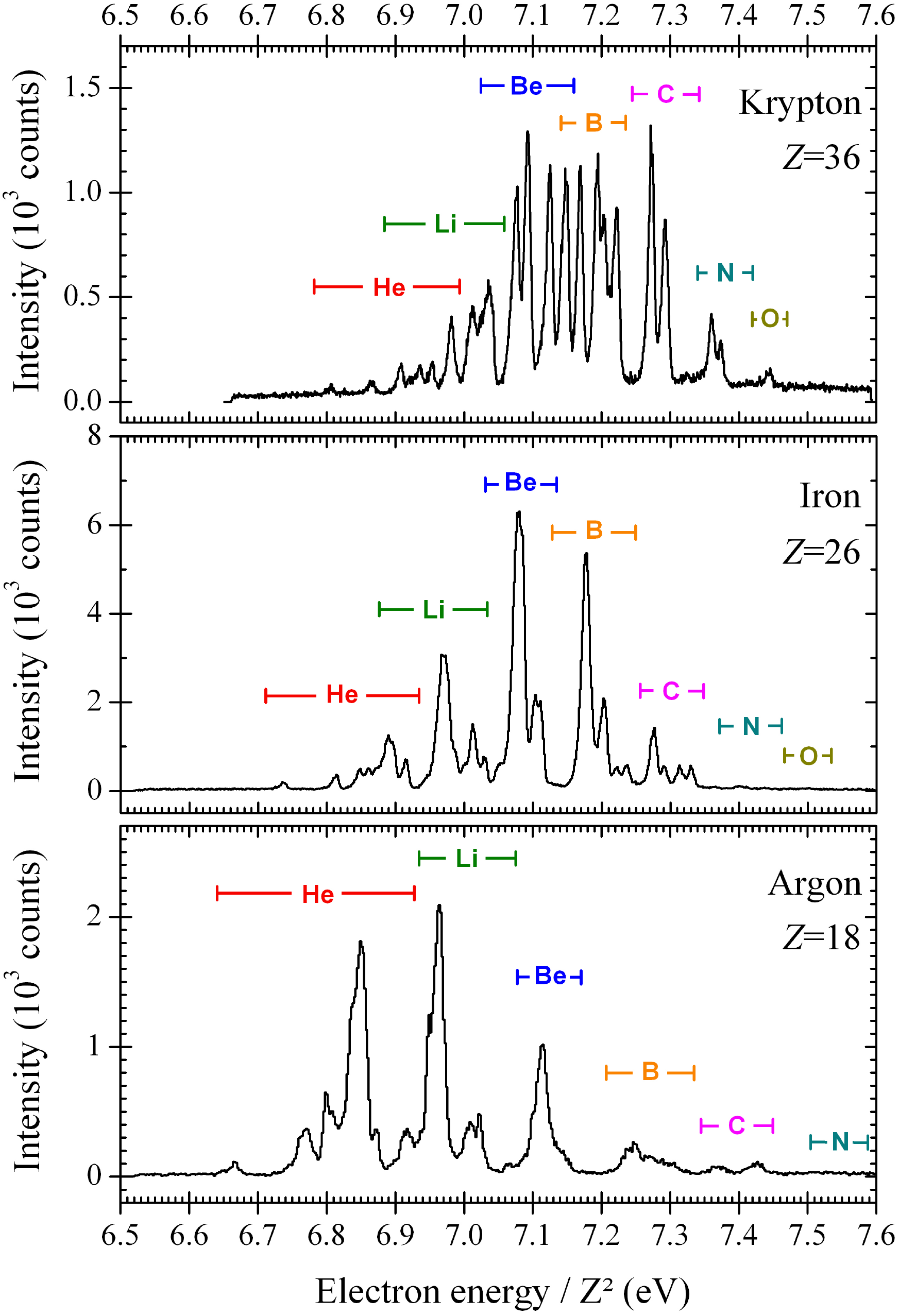}
\caption{\label{Kr-Fe-Ar-norm} (Color online) Electronic recombination spectra in the energy region of the $K-LL$ DR resonances and their higher orders
(especially TR) in highly charged Kr, Fe and Ar ions. The electron energy scales are multiplied by a factor $1/Z^2$.}
\end{figure}

We investigate the $K$-$LL$ energy region of resonant recombination into the He- to O-like charge states of
Ar, Fe and Kr ions in order to gain insight in the $Z$-behavior of resonance energies and strengths and systematic studies of higher-order processes. 
In an EBIT, different charge states of an element are stored simultaneously. By using a constant strong injection of neutrals, the charge state distribution
can be made sufficiently flat in order to make recombination resonances from He- to O-like ions appear together in the spectra. 
The high resolution allows for a good separation of the resonance groups for different ionic species, as indicated in Fig.~\ref{Kr-Fe-Ar-norm} by bars above each spectrum.

In this figure, we show the recombination spectra of the investigated ions in the specified energy region. 
We calibrate the electron energy scale by taking the predicted values of two well-separated resonance lines of He-like ions which have small theoretical
uncertainties. In the case of Ar it was possible to use two strong resonances from the $K$-$LL$ and $K$-$LM$ regions. 
In order to account for the $Z$-dependence of the resonance energies, the energy scales of the spectra are normalized by multiplying with a  factor $1/Z^{2}$ (Rydberg scaling).

The curves shown in Fig.~\ref{Kr-Fe-Ar-norm} 
display resonance energies, and relative resonance strengths, as well as the respective $Z$ dependences. 
In the normalized representation given, the $Z$-dependences of the resonance energies
(beyond the dominant $Z^{2}$-dependence) can be seen.

\begin{figure}[tb]
\includegraphics[width = \columnwidth]{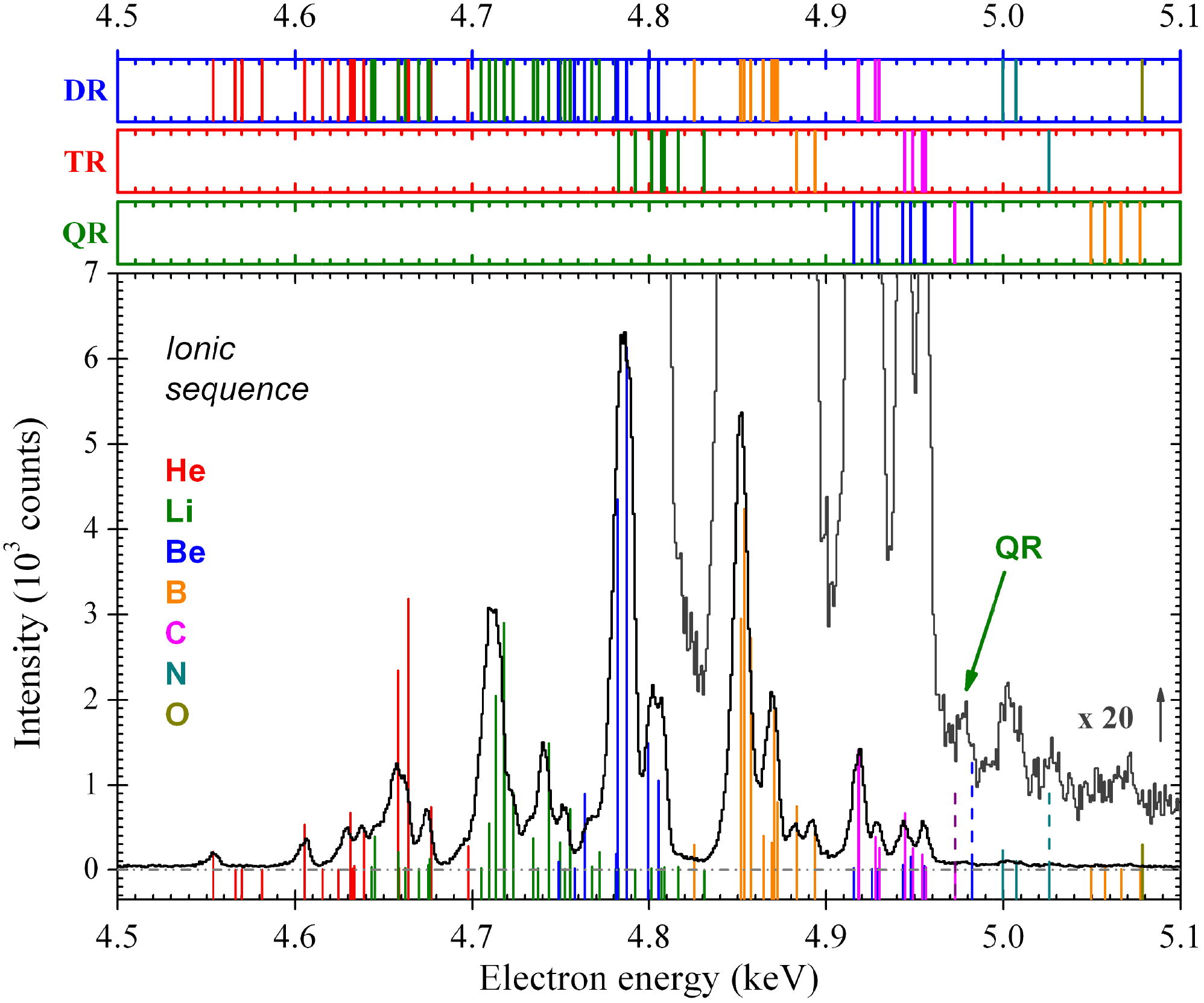}
\caption{\label{fe-kll} (Color online) Comparison of the Fe photorecombination spectrum with our theoretical predictions.
To show weak details, a 20-fold magnified spectrum is also displayed. On the top, the resonances
belonging to different recombination orders are indicated separately, and are color-coded according to their He-like to O-like charge states.}
\end{figure}

In Fig.~\ref{fe-kll}, we compare our recombination spectra for the astrophysically
relevant Fe ions with our theoretical predictions, where the abundances of the different charge state fractions have been adjusted
to match the measured ones. 

In general, a very good agreement both in the position of the resonances relative to those of the calibration resonances of He-like ions,
and in (relative) resonance strengths can be stated. This strengthens our confidence on the reliability of the calculations also for the
higher-order resonances, which we identify based on their predicted energy values.

Contributions from TR resonances can clearly be seen in Fig.~\ref{fe-kll} (as well as in Fig.~\ref{Kr-Fe-Ar-norm}) for initially B-like, C-like, and N-like ions
around 4.88, 4.95, and 5.04 keV, respectively. The DR spectrum for Li-like ions is especially rich and broad for $K$-$L$ excitation since not only the $2p$ levels
are accessible but also the $2s$ level has a vacancy. According to the calculations, TR resonances from Li-like ions are hidden under the strong DR lines of the
Be-like sequence.

Beyond the lowest-order $Z^2$-scaling of the binding energy, we have to consider the electron-electron interaction decreasing with $Z^{-1}$ in importance
relative to the central Coulomb force, and relativistic effects increasing rapidly with $Z^4$ for heavy ions. For the heavier species, the influence of the
latter is dominant. The range of ions studied here is characterized by a transition from the electron correlation-dominated range to one dominated by
relativistic effects, leading already here to a large fine-structure splitting. Hence, for the heavier system of Kr, and, to some extent also for Fe,
we find more lines resolved in the spectra. For the measurement with Ar ions, the resonances within one charge state overlap more strongly, and despite
the high experimental absolute resolution of only about 5~eV they can be barely resolved. In comparison to the transition energies, the relative resolution
is roughly equivalent to the absolute one of 13~eV seen in the case of Kr as the heaviest considered element. 

\begin{figure}[tb]
\includegraphics[width = \columnwidth]{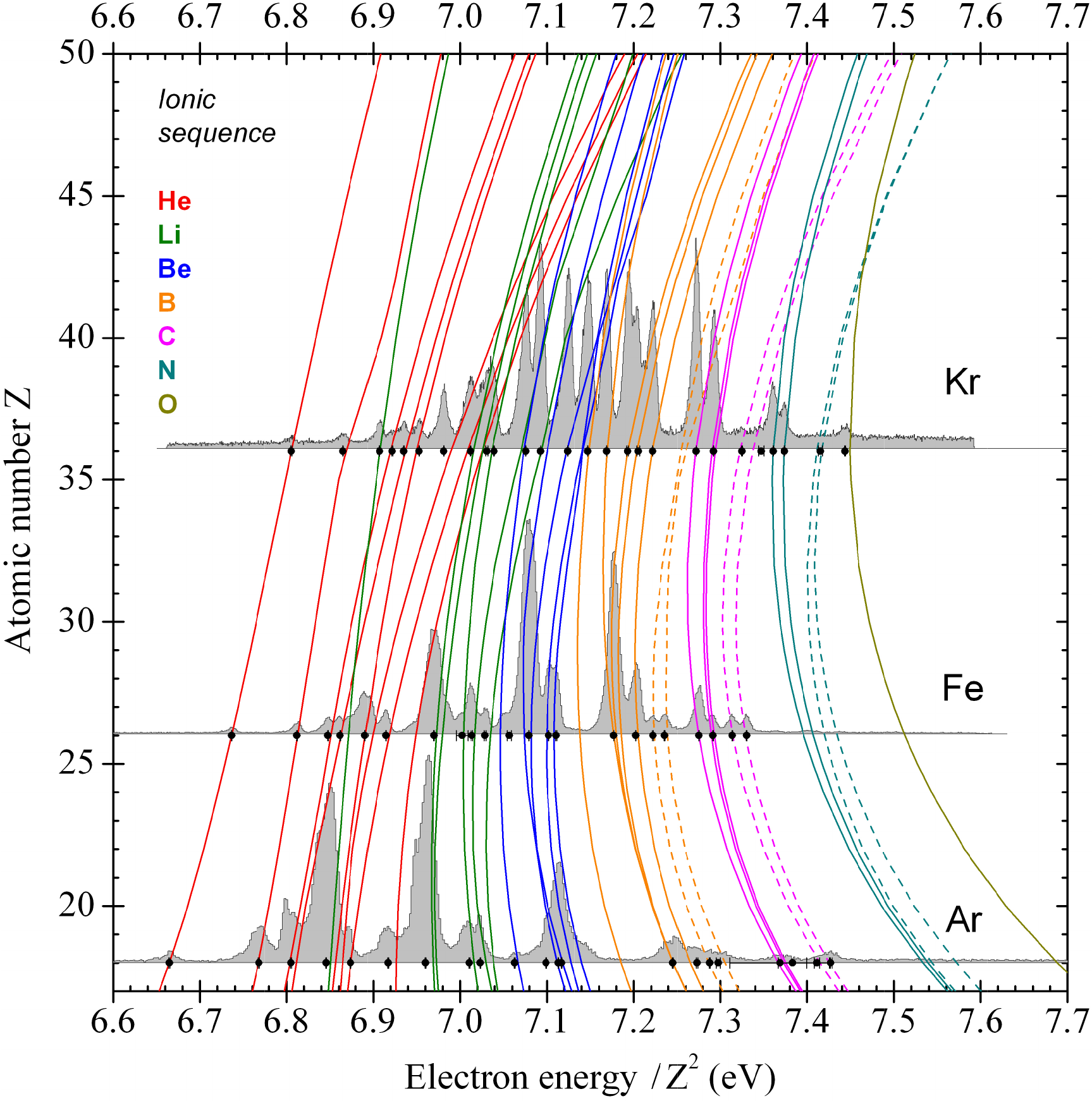}
\caption{\label{DR(Z)energies} (Color online) Photorecombination resonance energies for ions as a function of the atomic number $Z$; the resonance energies
are normalized by $Z^2$. The calculations are described in Section~\ref{theo}. Experimental values from the measured spectra for $Z$=18, 26, 36 (Ar, Fe, Kr, respectively)
appear as dots overlaid in gray. Trielectronic recombination resonances are marked by dashed lines.}
\end{figure}

In Fig.~\ref{DR(Z)energies} we show the $Z$ dependence of the relative resonance energies 
for the strongest recombination lines of initially He-like to O-like ions. 
The curves normalized by $Z^2$ are based on our theoretical values including electron-electron correlations. 
These spectra show clearly the transition region, from the strong electron-electron interaction at low $Z$ to the increased relativistic influence at higher $Z$. 
These trends are confirmed experimentally, as spectra for Ar, Fe and Kr overlaid to the graph in  Fig.~\ref{DR(Z)energies} demonstrate. 

We now discuss the resonance strengths of C-like ions, which among all isoelectronic sequences with an open $L$ shell are expected to have the strongest TR contributions.
The occupation rules -- partially also in combination with parity rules (cf. Section \ref{theo}) -- allow for several resonances. 
With initially four vacancies in the open $L$-shell and with two $2s$ as well as $2p$ electrons, many possible excitation as well as relaxation channels for TR processes appear. In contrast, the B-like sequence has a smaller number of excitable electrons, and N-like ions contain less vacancies into which excitation can occur. For the Li-like sequence there is only one $2s$ electron left for the additional intra-shell excitation in TR. These differences in the recombination strengths of TR in different ionic sequences
are also illustrated below  in Fig.~\ref{TR/DR-strength}.

\begin{figure}[tb]
\includegraphics[width = \columnwidth]{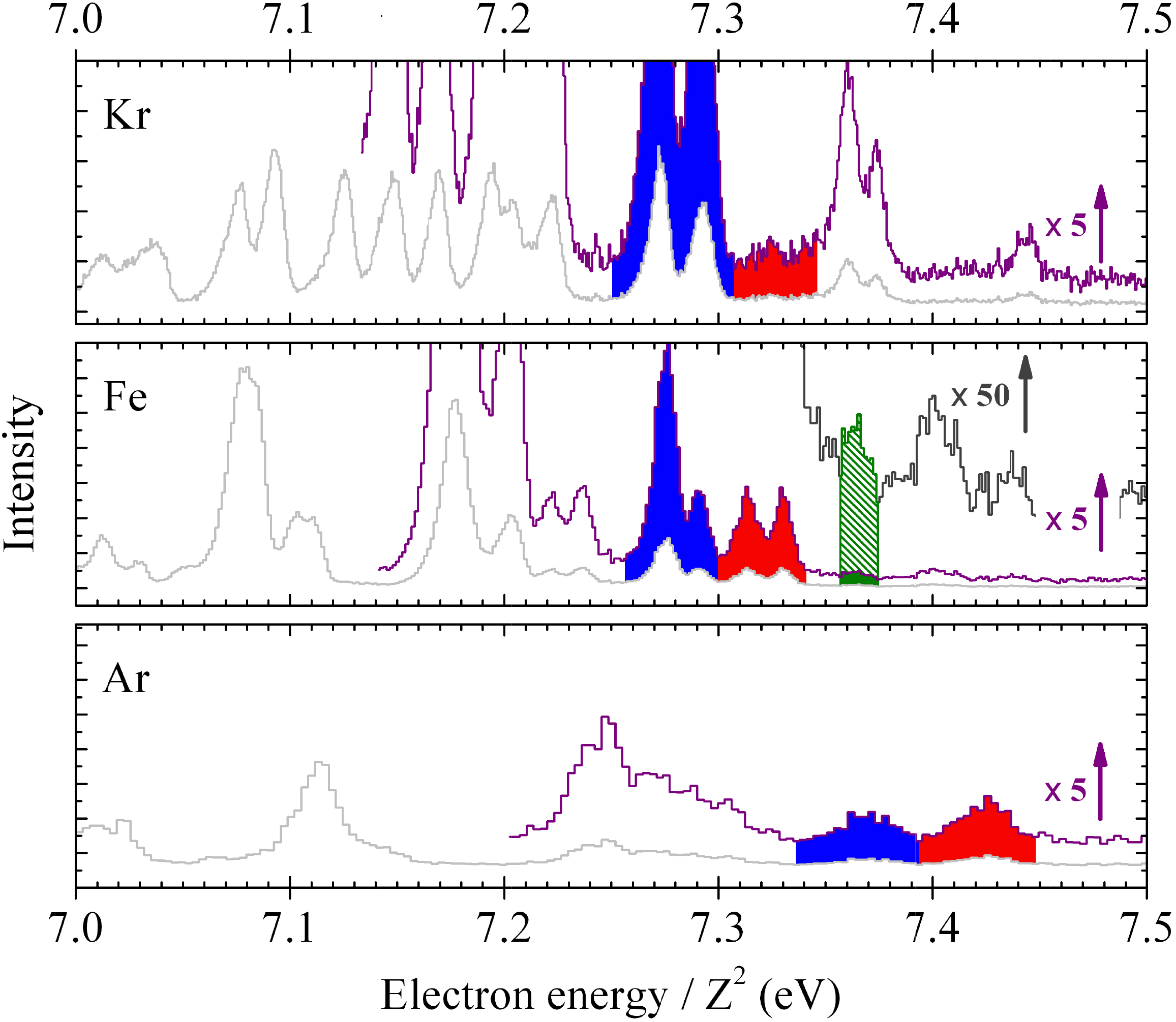}
\caption{\label{ar_fe_kr_blowup} (Color online) Partial magnified spectrum showing photorecombination resonances of Be-like to N-like ions of
Ar, Fe, and Kr. DR, TR and QR processes are marked for C-like ions.
}
\end{figure}

The region of interest of the recombination spectra for the C-like isoelectronic sequence is enlarged in Fig.~\ref{ar_fe_kr_blowup}. 
The resonance areas for DR and TR are indicated in full blue and red, respectively. 
For Kr the spectrum is equivalent 
to our data from \cite{Beilmann09}, with the two prominent DR double lines for C-like ions (at 7.27 and 7.28 eV/$Z^2$) followed by the double DR line for N-like Kr ions (at 7.36 and 7.37 eV/$Z^2$). 
In between (at around 7.32 eV/$Z^2$) we find the faint structure for C-like TR process contributing in total only 6$\%$ to all the resonant recombination processes in this C-like case. 

Proceeding down in $Z$ to the Fe ions, the importance of electron-electron interaction increases and, correspondingly, TR can be expected to be more important there. 
For C-like Fe ions we find the $K$-$LL$ DR double line at 7.28 and 7.30 eV/$Z^2$ and the $KL$-$LLL$ TR
double line at 7.32 and 7.34 eV/$Z^2$; however, the relative TR contribution increases already dramatically to about 50$\%$ of the
DR contribution~\cite{Beilmann11a}. 
Continuing further down in $Z$ to Ar, the doublets for C-like ions are not any more resolved
experimentally due to the reduced fine structure splitting of the contributing different $j$ components --
see the broad $K$-$LL$ DR line at 7.37 eV/$Z^2$. 
The stronger line around 7.42 eV/$Z^2$ corresponds to the higher-order $KL$-$LLL$ TR process of C-like Ar ions. 
Here, the  TR process overwhelms already the corresponding first-order $K$-$LL$ DR by a factor of about 1.5. 
This strong effect shows how higher-order electron-electron correlation can surpass the (intuitively stronger) first-order process~\cite{Beilmann11a}. 

\begin{figure}[tb]
\includegraphics[width = 0.9\columnwidth]{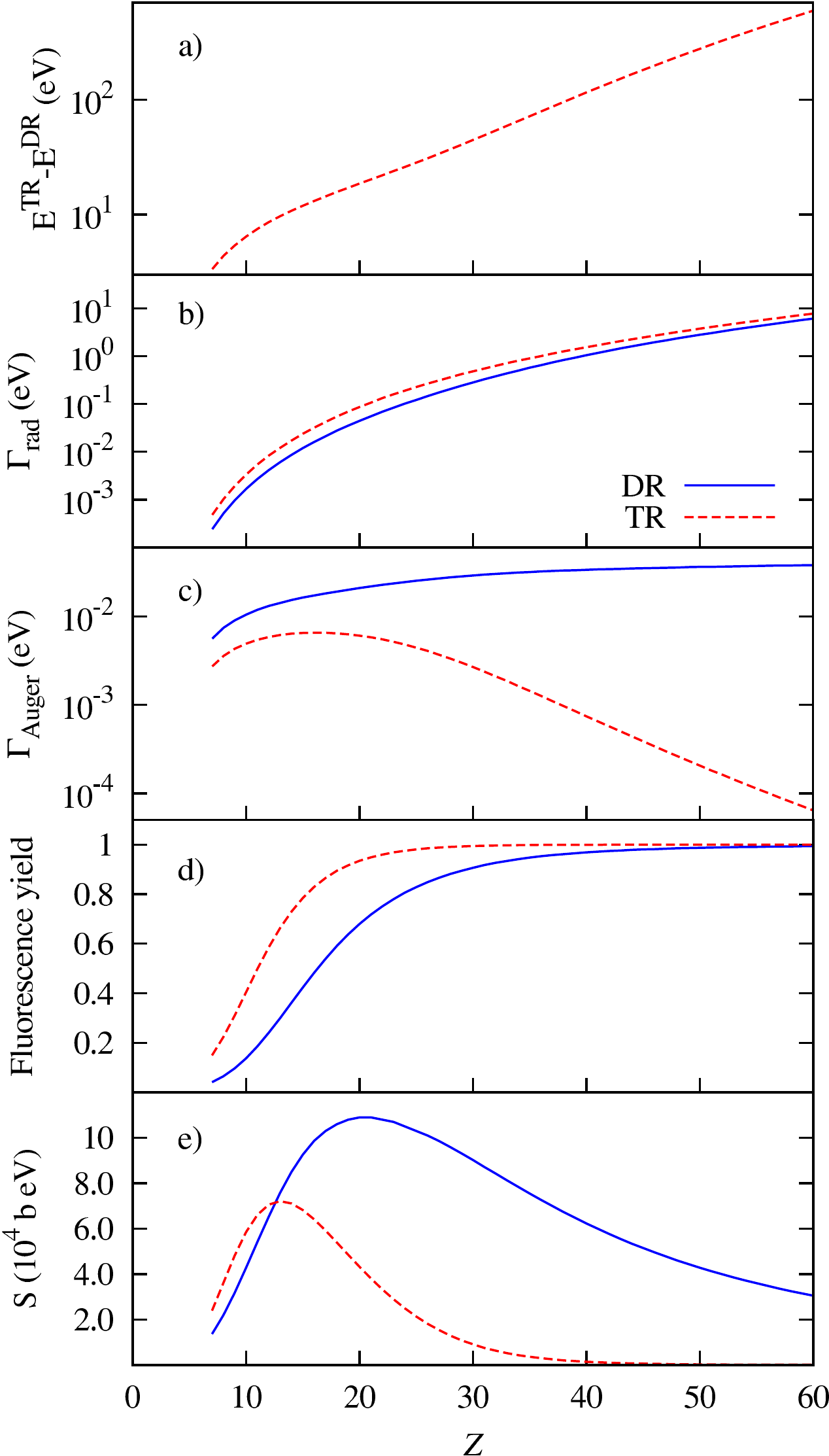}
\caption{\label{fig_scaling_theo_5half}
(Color online) Theoretical scaling of different atomic quantities contributing to the DR and TR
resonant recombination resonance strength as a function of $Z$, for $J=5/2$ resonances in initially C-like ions.
(a) difference of TR and DR resonance energies; (b) radiative decay rates; (c) Auger decay rates,
(d) fluorescence yields, and (e) resonance strengths. The quantities are given for the case of the
$1s 2s^2 2p_{1/2}^2 (2p_{3/2}^2)_2$ $J=5/2$ DR resonance (full blue line) and the $(1s 2s^2 2p_{1/2})_1 2p_{3/2}^3$ $J=5/2$ TR resonance (dashed red line).
}
\end{figure}

In order to understand this unexpected increase in the relative importance of the higher-order process of TR at low $Z$ compared to DR,
one has to consider the absolute resonance strength. 
Following the general formula~(\ref{eq:strength}), it can be described as the product of the Auger rate responsible for the population of the excited intermediate state and the fluorescence yield determining its decay to the final state. 
The fluorescence yield is determined by the ratio of the radiative decay rate to the total decay rate,
which is the sum over all possible radiative and Auger channels. 

\begin{figure}[tb]
\includegraphics[width = 0.9\columnwidth]{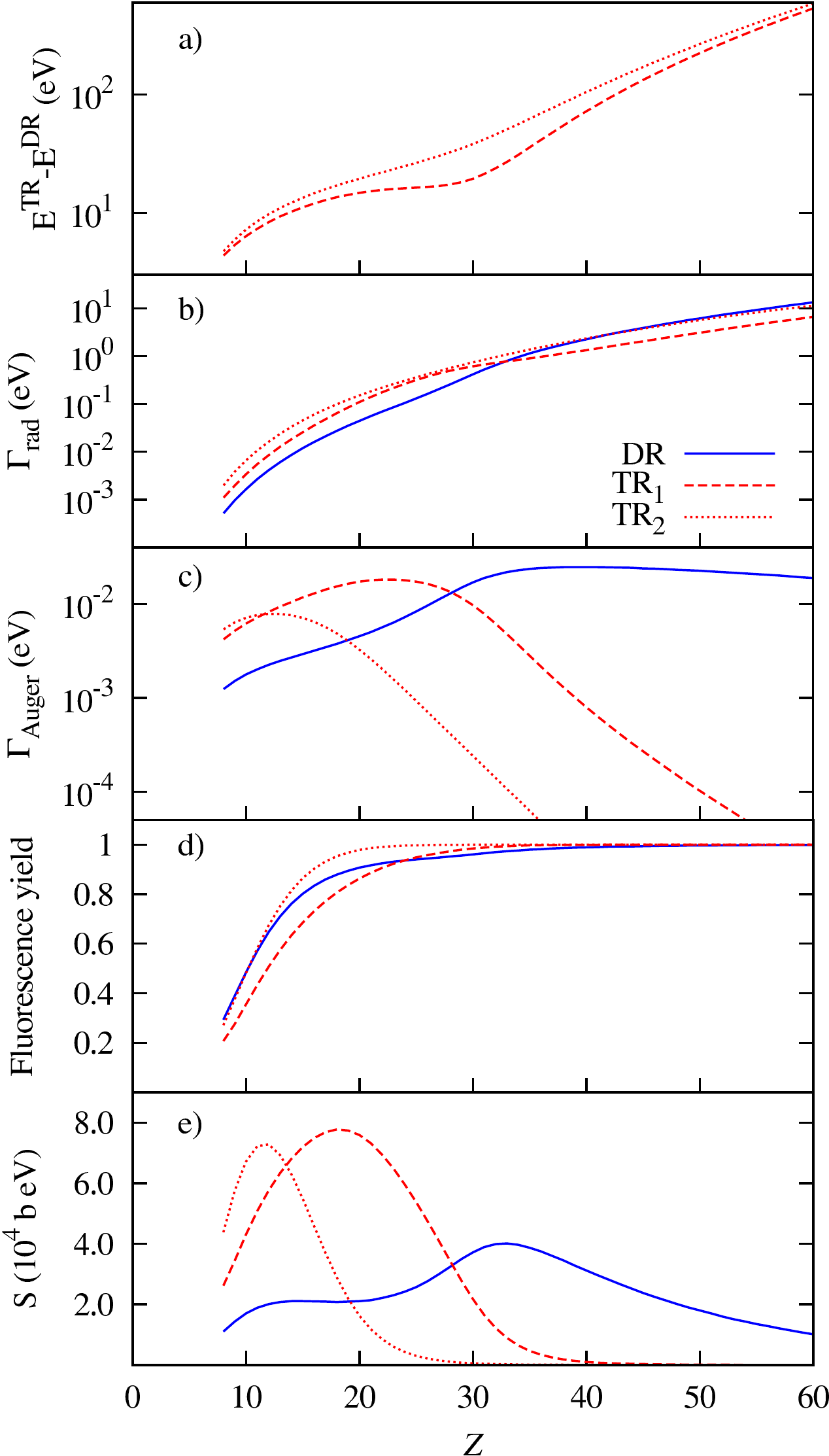}
\caption{\label{fig_scaling_theo} (Color online) Same as Fig.~\ref{fig_scaling_theo_5half}, but for $J=3/2$
resonances in initially C-like ions. These quantities are given for the case of the $1s 2s^2 2p_{1/2}^2 (2p_{3/2}^2)_2$ $J=3/2$ DR
resonance (straight blue line) and the $(1s 2s^2 2p_{1/2})_1 2p_{3/2}^3$ (TR$_1$, dashed red line) and
$(1s 2s^2 2p_{1/2})_0 2p_{3/2}^3$ (TR$_2$, dotted red line) $J=3/2$ TR resonances.}
\end{figure}

The calculated $Z$-scalings for the contributing parameters for selected DR and TR resonances in initially C-like ions are sketched in
Figs.~\ref{fig_scaling_theo_5half} and \ref{fig_scaling_theo}. Resonances with total angular momenta $J=5/2$ and $J=3/2$ have been
selected for demonstration; note that DR and TR states may mix with each other within a set of levels possessing the same $J$.
The radiative rates (b) increase generally with $\Gamma^{r} \propto Z^4$
(see, e.~g.~\cite{Burgess64} or the overview in \cite{Mokler78}), whereas the normal (two-electron) Auger rates (c), relevant for DR,
are almost $Z$-independent, $\Gamma^{a,2e} \propto Z^{0}$. Relativistic wave function and Breit interaction corrections
introduce terms scaling with even powers of $Z$ at higher charge numbers~\cite{Zimmerer91}; however, these are not of
relevance for the light and medium-heavy elements where higher-order recombination processes are observable.

For high $Z$ values, in the case of DR, the fluorescence yield is dominated by the radiative decay and thus it approaches unity, whereas
for small $Z$ it is governed by the Auger rate, see Fig.~\ref{fig_scaling_theo_5half}\,d. Following Eq.~(\ref{eq:strength}) and including
the phase-space factor $1/p^2 \approx 1/(2E)$, this yields the following scaling law for the strength~\cite{Watanabe02,Kavanagh10}:
\begin{equation}
\label{eq_DR_strength}
S^{\text{DR}} \propto \frac{1}{Z^{2}} \frac{Z^{4}Z^{0}}{a_{1}Z^{4} + a_{2}Z^{0}} = \frac{1}{a_{1}Z^{2} + a_{2}Z^{-2}} \,.
\end{equation}
Thus, in a simple nonrelativistic approximation, the DR strength is roughly proportional to $Z^2$ at low atomic numbers, decreases for high $Z$ with $1/Z^2$,
and it is almost constant in the mid-$Z$ region.

In a similar way, scaling laws may be derived for the higher-order resonances. The $Z$-dependence of the contributing parameters
for the TR processes are shown as dashed lines in Figs.~\ref{fig_scaling_theo_5half} and \ref{fig_scaling_theo}. The Auger rates causing the
higher-order resonances show a somewhat different behavior. The perturbative factor in the first line of
Eq.~(\ref{eq:tr-pert}) introduces an additional scaling factor of $1/Z$ to the transition amplitude of TR, therefore,
the three-electron Auger rates (and the TR capture rates) decrease approximately as $\Gamma^{a,3e} \propto Z^{-2}$ at high
$Z$ values. This three-electron Auger rate has also to be taken into consideration in the calculation of the total decay rate,
and thus the $Z$-scaling of the TR resonance strength follows the parameterization:
\begin{equation}
\label{eq_TR_strength}
S^{\text{TR}} \propto \frac{1}{Z^{2}} \frac{Z^{4}Z^{-2}}{b_{1}Z^{4} + b_{2}Z^{0} + b_{3}Z^{-2}} = \frac{1}{b_{1}Z^{4} + b_{2} + b_{3}Z^{-2}} \,.
\end{equation}
At higher atomic numbers, the TR resonance strength decreases as $\propto Z^{-4}$, i.~e., significantly faster than the
DR strength, therefore, the TR process can hardly be observed in heavier ions.

Independently on that feature, at low $Z$ values, the second-order Auger rate responsible for TR is comparable to
or larger than the first-order Auger rate corresponding to DR (see the case with $J=3/2$ in Fig.~\ref{fig_scaling_theo}\,c). The rate for radiative
decay following trielectronic capture may also be larger than the corresponding radiative decay rate for DR (Fig.~\ref{fig_scaling_theo}\,b),
resulting in a TR resonance stronger than DR.
In some cases, e.g. in $J=5/2$ resonances in initially C-like ions, the three-electron Auger rate may even be weaker
than the two-electron Auger rate (see Fig.~\ref{fig_scaling_theo_5half}\,c), yet the larger radiative rate overcompensates
this and yields a higher strength for TR than for DR.

\begin{figure}[tb]
\includegraphics[width = \columnwidth]{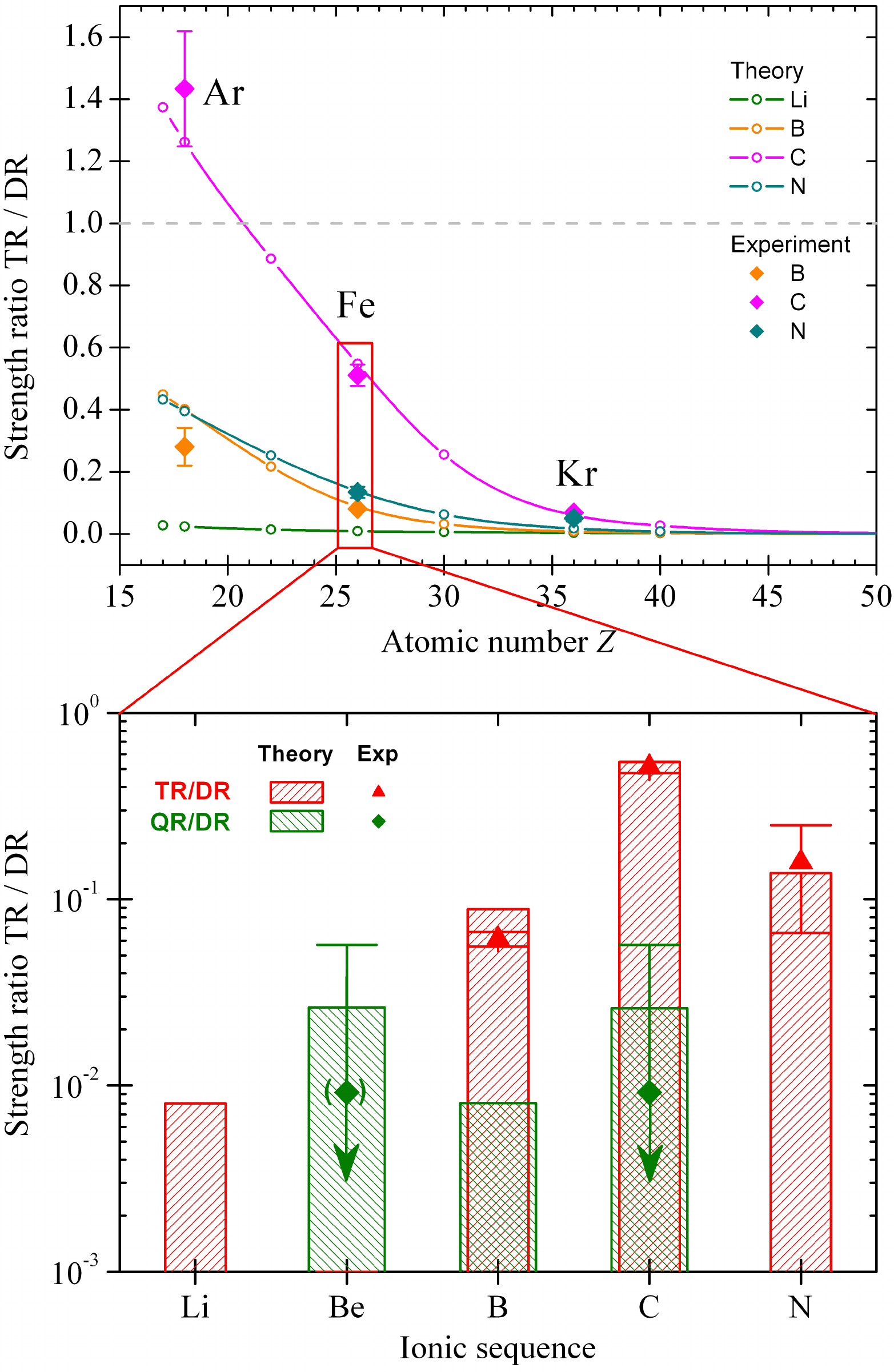}
\caption{\label{TR/DR-strength} (Color online) Higher-order TR contributions relative to the corresponding DR resonance strengths
as functions of the atomic number $Z$ for Li-like to N-like ionic species. The calculations are compared with the experimental results.
(Full gray points are calculated ab initio, dashed curves show interpolations; experimental data points show error bars.)
For Fe ions, the strength ratios TR/DR and QR/DR for different charge states are given in the bottom graph.}
\end{figure}

In the present experiments the total absolute resonance strengths are not accessible directly. However, the calculated ratio of the strength of TR
and the corresponding DR process within a single ionic sequence can be compared with the measurements, see the top panel of Fig.~\ref{TR/DR-strength}.
As discussed above, for low $Z$, the importance of TR increases and for C-like ions TR dominates over the DR process for $Z < 20$. 
The sequence for B- and N-like ions shows a similar behavior, but for these charge states TR does not have a dominant influence on the total recombination yield. 
For Li-like ions, TR contributes only little to the recombination because of the limited number of excitation channels: only one electron
is available in the $L$ shell to be excited simultaneously with a $K$-shell electron.

One can state for the TR/DR strength ratio an excellent agreement between our experimental findings and our calculations.
This is not only true for the $Z$-dependence of the strength ratio, but holds also for the dependence on the ionic sequence.
In the bottom panel of Fig~\ref{TR/DR-strength} we display for Li- to N-like Fe ions the corresponding strength ratio TR/DR including also the
ratio QR/DR.

For Fe ions, the nuclear charge $Z$ is low enough to show sizable contributions of higher-order resonances
and high enough to ensure a good resolution of the resonance spectrum. 
The QR/DR strength ratio seems also be confirmed within the experimental uncertainty by the calculations.
Although the statistics for QR are insufficient for a precise determination of resonance energies, the spectrum shown in Figs.~\ref{fe-kll} and \ref{ar_fe_kr_blowup} shows an obvious indication of QR resonances. Our calculations and the experimental findings agree convincingly well (see the correspondingly marked resonance in Fig.~\ref{fe-kll}). 
The faint structures seen around an electron energy of 4.980\,keV correspond thus to QR in both C- and Be-like Fe ions. 

An approximate scaling law for QR may be derived in an analogous manner to that found for TR, using similar considerations for
the four-electron Auger rate ($\Gamma^{a,4e} \propto Z^{-4}$):
\begin{eqnarray}
S^{\rm QR} &\propto& 1/Z^2 \frac{Z^4 Z^{-4}}{c_1 Z^4 + c_2 Z^0 + c_3 Z^{-2} + c_4 Z^{-4}} \nonumber \\
&=& \frac{1}{c_1 Z^6 + c_2 Z^2 + c_3 + c_4 Z^{-2}} \,.
\end{eqnarray}
Here, we have taken into account that autoionizing states formed by quadruelectronic capture may decay by two-, three- or four-electron Auger processes.
The QR resonance strength diminishes for high $Z$ even faster than the strength of TR, namely, as $\propto Z^{-6}$.
However, its experimental signatures have become measurable, and a systematic study of the interesting isoelectronic sequences as a function of $Z$ could provide better insight into the complex mutual interactions of same-shell electrons,
beyond what is possible by studying the single neutrals giving name to those sequences.

\section{\label{conclusions} Conclusions}

Unexpected large contributions of higher-order electronic recombination processes have been found for inter-shell
$K$-$L$ excitation. For these processes, high momenta are exchanged during recombination and hence,
large contributions from higher orders are generally not expected. As shown, the second-order resonant electron
recombination process, TR, may even overwhelm the strength of the first-order DR process, as observed in the case
of C-like Ar, cf. Fig.~\ref{TR/DR-strength}. In most cases, the second-order TR process contributes about
10\% to the total electronic recombination cross section, and its relative contribution decreases with increasing $Z$.
Even the third-order process, QR, was clearly identified, contributing typically a few \% of DR.

These novel findings on the importance of higher-order processes in resonant inter-shell electronic recombination
were made possible by enhanced energy resolution and good experimental statistics, both points being achieved by
evaporative axial cooling of the trapped ions.

Beyond the obvious $Z$ trends seen in the resonance energies and caused by the balance between electron correlation and
relativistic effects, strikingly apparent higher-order electron-electron correlation effects are found at lower
atomic numbers. Here, the strength of higher-order recombination processes can surpass that of the corresponding
first-order DR process~\cite{Beilmann11a}.

Since the {\it inter-shell} $K$-$LL$ photorecombination process is a far more efficient cooling mechanism in high-temperature plasmas
than {\it intra-shell} processes, due to the much higher energy of the emitted photons, a good knowledge of its energy-dependent cross section is essential.
Therefore, hitherto neglected contributions from higher-order effects to electron recombination, and especially of the second-order, the trielectronic channel,
have to be taken into account. Plasma simulations used for modeling fusion devices or astrophysical matter have to include them
for improved accuracy. The presence of sizable TR resonances has implications for the opacity and energy transfer inside the radiative zone of stars,
and for the temperature and ionization equilibrium of those and other astrophysical plasmas. Models not considering these higher-order
effects in plasmas containing light-$Z$ species can therefore not describe correctly the status in a plasma.
Moreover, for the time-reversed processes of resonant photoionization~\cite{Simon10a}, i.~e., the resonant excitation to an autoionizing
intermediate state, also those higher-order processes may be of equal importance. Even beyond TR, quadruelectronic recombination has been
observed for different ionic species.

Studying photorecombination in all orders of complexity, from DR to TR and QR, can become a fruitful approach for
understanding more complex multi-electron processes such as shake-up and shake-off in outer shells, especially for
heavier species~\cite{Aksela96a,Uiberacker07}. Arguably, even higher recombination orders involving more than four
simultaneous electronic excitations, which are more likely to be expected in recombination of open $M$-shell ions
having many equivalent electrons and vacancies energetically closely spaced, should be investigated to learn about
the stepwise buildup of a many-electron wave function. The role of relativistic corrections to the Coulomb electron interaction,
i.e., the Breit interaction~\cite{Zimmerer91,Fritzsche2009}, in such higher-order processes is also an unexplored field. With the ion charge as
a variable parameter and a constant number of electrons, isoelectronic sequences in highly charged ions are perfectly
suited for this type of investigations. Furthermore, such a multi-electron behavior may also be manifested in resonant
processes occurring due to the presence of neighboring atoms or ions in dense plasmas or molecules \cite{MuellerC10,Voitkiv10}.

\begin{acknowledgments}

The work of ZH was supported by the Alliance Program of the Helmholtz Association (HA216/EMMI).

\end{acknowledgments}


\begin{thebibliography}{41}
\expandafter\ifx\csname natexlab\endcsname\relax\def\natexlab#1{#1}\fi
\expandafter\ifx\csname bibnamefont\endcsname\relax
  \def\bibnamefont#1{#1}\fi
\expandafter\ifx\csname bibfnamefont\endcsname\relax
  \def\bibfnamefont#1{#1}\fi
\expandafter\ifx\csname citenamefont\endcsname\relax
  \def\citenamefont#1{#1}\fi
\expandafter\ifx\csname url\endcsname\relax
  \def\url#1{\texttt{#1}}\fi
\expandafter\ifx\csname urlprefix\endcsname\relax\def\urlprefix{URL }\fi
\providecommand{\bibinfo}[2]{#2}
\providecommand{\eprint}[2][]{\url{#2}}

\bibitem[{\citenamefont{Massey and Bates}(1942)}]{Massey42}
\bibinfo{author}{\bibfnamefont{H.~S.~W.} \bibnamefont{Massey}}
  \bibnamefont{and} \bibinfo{author}{\bibfnamefont{D.~R.} \bibnamefont{Bates}},
  \bibinfo{journal}{Rep. Prog. Phys.} \textbf{\bibinfo{volume}{9}},
  \bibinfo{pages}{62} (\bibinfo{year}{1942}).

\bibitem[{\citenamefont{Burgess}(1964)}]{Burgess64}
\bibinfo{author}{\bibfnamefont{A.}~\bibnamefont{Burgess}},
  \bibinfo{journal}{Astrophys. J.} \textbf{\bibinfo{volume}{139}},
  \bibinfo{pages}{776} (\bibinfo{year}{1964}).

\bibitem[{\citenamefont{M\"uller}(2008)}]{Mueller08}
\bibinfo{author}{\bibfnamefont{A.}~\bibnamefont{M\"uller}},
  \bibinfo{journal}{Adv. At., Mol., Opt. Phys.} \textbf{\bibinfo{volume}{55}},
  \bibinfo{pages}{293} (\bibinfo{year}{2008}).

\bibitem[{\citenamefont{Flambaum et~al.}(2002)\citenamefont{Flambaum,
  Gribakina, Gribakin, and Harabati}}]{Flambaum02}
\bibinfo{author}{\bibfnamefont{V.~V.} \bibnamefont{Flambaum}},
  \bibinfo{author}{\bibfnamefont{A.~A.} \bibnamefont{Gribakina}},
  \bibinfo{author}{\bibfnamefont{G.~F.} \bibnamefont{Gribakin}},
  \bibnamefont{and} \bibinfo{author}{\bibfnamefont{C.}~\bibnamefont{Harabati}},
  \bibinfo{journal}{Phys. Rev. A} \textbf{\bibinfo{volume}{66}},
  \bibinfo{pages}{012713} (\bibinfo{year}{2002}).

\bibitem[{\citenamefont{{Gonz\'{a}lez Mart\'{\i}nez}
  et~al.}(2005)\citenamefont{{Gonz\'{a}lez Mart\'{\i}nez}, {Crespo
  L\'{o}pez-Urrutia}, Braun, Brenner, Bruhns, Lapierre, Mironov, {Soria Orts},
  Tawara, Trinczek et~al.}}]{Gonzalez05}
\bibinfo{author}{\bibfnamefont{A.~J.} \bibnamefont{{Gonz\'{a}lez
  Mart\'{\i}nez}}}, \bibinfo{author}{\bibfnamefont{J.~R.} \bibnamefont{{Crespo
  L\'{o}pez-Urrutia}}},
  \bibinfo{author}{\bibfnamefont{J.}~\bibnamefont{Braun}},
  \bibinfo{author}{\bibfnamefont{G.}~\bibnamefont{Brenner}},
  \bibinfo{author}{\bibfnamefont{H.}~\bibnamefont{Bruhns}},
  \bibinfo{author}{\bibfnamefont{A.}~\bibnamefont{Lapierre}},
  \bibinfo{author}{\bibfnamefont{V.}~\bibnamefont{Mironov}},
  \bibinfo{author}{\bibfnamefont{R.}~\bibnamefont{{Soria Orts}}},
  \bibinfo{author}{\bibfnamefont{H.}~\bibnamefont{Tawara}},
  \bibinfo{author}{\bibfnamefont{M.}~\bibnamefont{Trinczek}},
  \bibnamefont{et~al.}, \bibinfo{journal}{Phys. Rev. Lett.}
  \textbf{\bibinfo{volume}{94}}, \bibinfo{pages}{203201}
  (\bibinfo{year}{2005}).

\bibitem[{\citenamefont{Harman et~al.}(2006)\citenamefont{Harman, Tupitsyn,
  Artemyev, Jentschura, Keitel, {Crespo L\'{o}pez-Urrutia}, {Gonz\'{a}lez
  Mart\'inez}, Tawara, and Ullrich}}]{Harman06}
\bibinfo{author}{\bibfnamefont{Z.}~\bibnamefont{Harman}},
  \bibinfo{author}{\bibfnamefont{I.~I.} \bibnamefont{Tupitsyn}},
  \bibinfo{author}{\bibfnamefont{A.~N.} \bibnamefont{Artemyev}},
  \bibinfo{author}{\bibfnamefont{U.~D.} \bibnamefont{Jentschura}},
  \bibinfo{author}{\bibfnamefont{C.~H.} \bibnamefont{Keitel}},
  \bibinfo{author}{\bibfnamefont{J.~R.} \bibnamefont{{Crespo
  L\'{o}pez-Urrutia}}}, \bibinfo{author}{\bibfnamefont{A.~J.}
  \bibnamefont{{Gonz\'{a}lez Mart\'inez}}},
  \bibinfo{author}{\bibfnamefont{H.}~\bibnamefont{Tawara}}, \bibnamefont{and}
  \bibinfo{author}{\bibfnamefont{J.}~\bibnamefont{Ullrich}},
  \bibinfo{journal}{Phys. Rev. A} \textbf{\bibinfo{volume}{73}},
  \bibinfo{pages}{052711} (\bibinfo{year}{2006}).

\bibitem[{\citenamefont{Brandau et~al.}(2008)\citenamefont{Brandau, Kozhuharov,
  Harman, M\"{u}ller, Schippers, Kozhedub, Bernhardt, B\"{o}hm, Jacobi, Schmidt
  et~al.}}]{Brandau08}
\bibinfo{author}{\bibfnamefont{C.}~\bibnamefont{Brandau}},
  \bibinfo{author}{\bibfnamefont{C.}~\bibnamefont{Kozhuharov}},
  \bibinfo{author}{\bibfnamefont{Z.}~\bibnamefont{Harman}},
  \bibinfo{author}{\bibfnamefont{A.}~\bibnamefont{M\"{u}ller}},
  \bibinfo{author}{\bibfnamefont{S.}~\bibnamefont{Schippers}},
  \bibinfo{author}{\bibfnamefont{Y.~S.} \bibnamefont{Kozhedub}},
  \bibinfo{author}{\bibfnamefont{D.}~\bibnamefont{Bernhardt}},
  \bibinfo{author}{\bibfnamefont{S.}~\bibnamefont{B\"{o}hm}},
  \bibinfo{author}{\bibfnamefont{J.}~\bibnamefont{Jacobi}},
  \bibinfo{author}{\bibfnamefont{E.~W.} \bibnamefont{Schmidt}},
  \bibnamefont{et~al.}, \bibinfo{journal}{Phys. Rev. Lett.}
  \textbf{\bibinfo{volume}{100}}, \bibinfo{pages}{073201}
  (\bibinfo{year}{2008}).

\bibitem[{\citenamefont{Schippers}(2009)}]{Schippers09}
\bibinfo{author}{\bibfnamefont{S.}~\bibnamefont{Schippers}},
  \bibinfo{journal}{J. Phys.: Conf. Ser.} \textbf{\bibinfo{volume}{163}},
  \bibinfo{pages}{012001} (\bibinfo{year}{2009}).

\bibitem[{\citenamefont{Knapp et~al.}(1995)\citenamefont{Knapp, Beiersdorfer,
  Chen, Scofield, and Schneider}}]{Knapp95}
\bibinfo{author}{\bibfnamefont{D.~A.} \bibnamefont{Knapp}},
  \bibinfo{author}{\bibfnamefont{P.}~\bibnamefont{Beiersdorfer}},
  \bibinfo{author}{\bibfnamefont{M.~H.} \bibnamefont{Chen}},
  \bibinfo{author}{\bibfnamefont{J.~H.} \bibnamefont{Scofield}},
  \bibnamefont{and}
  \bibinfo{author}{\bibfnamefont{D.}~\bibnamefont{Schneider}},
  \bibinfo{journal}{Phys. Rev. Lett.} \textbf{\bibinfo{volume}{74}},
  \bibinfo{pages}{54} (\bibinfo{year}{1995}).

\bibitem[{\citenamefont{Beilmann et~al.}(2009)\citenamefont{Beilmann,
  Postavaru, Arntzen, Ginzel, Keitel, M\"{a}ckel, Mokler, Simon, Tawara,
  Tupitsyn et~al.}}]{Beilmann09}
\bibinfo{author}{\bibfnamefont{C.}~\bibnamefont{Beilmann}},
  \bibinfo{author}{\bibfnamefont{O.}~\bibnamefont{Postavaru}},
  \bibinfo{author}{\bibfnamefont{L.~H.} \bibnamefont{Arntzen}},
  \bibinfo{author}{\bibfnamefont{R.}~\bibnamefont{Ginzel}},
  \bibinfo{author}{\bibfnamefont{C.~H.} \bibnamefont{Keitel}},
  \bibinfo{author}{\bibfnamefont{V.}~\bibnamefont{M\"{a}ckel}},
  \bibinfo{author}{\bibfnamefont{P.~H.} \bibnamefont{Mokler}},
  \bibinfo{author}{\bibfnamefont{M.~C.} \bibnamefont{Simon}},
  \bibinfo{author}{\bibfnamefont{H.}~\bibnamefont{Tawara}},
  \bibinfo{author}{\bibfnamefont{I.~I.} \bibnamefont{Tupitsyn}},
  \bibnamefont{et~al.}, \bibinfo{journal}{Phys. Rev. A}
  \textbf{\bibinfo{volume}{80}}, \bibinfo{pages}{050702}
  (\bibinfo{year}{2009}).

\bibitem[{\citenamefont{Beilmann et~al.}(2011)\citenamefont{Beilmann, Mokler,
  Bernitt, Keitel, Ullrich, {Crespo L\'{o}pez-Urrutia}, and
  Harman}}]{Beilmann11a}
\bibinfo{author}{\bibfnamefont{C.}~\bibnamefont{Beilmann}},
  \bibinfo{author}{\bibfnamefont{P.~H.} \bibnamefont{Mokler}},
  \bibinfo{author}{\bibfnamefont{S.}~\bibnamefont{Bernitt}},
  \bibinfo{author}{\bibfnamefont{C.~H.} \bibnamefont{Keitel}},
  \bibinfo{author}{\bibfnamefont{J.}~\bibnamefont{Ullrich}},
  \bibinfo{author}{\bibfnamefont{J.~R.} \bibnamefont{{Crespo
  L\'{o}pez-Urrutia}}}, \bibnamefont{and}
  \bibinfo{author}{\bibfnamefont{Z.}~\bibnamefont{Harman}},
  \bibinfo{journal}{Phys. Rev. Lett.} \textbf{\bibinfo{volume}{107}},
  \bibinfo{pages}{143201} (\bibinfo{year}{2011}).

\bibitem[{\citenamefont{Schnell et~al.}(2003)\citenamefont{Schnell, Gwinner,
  Badnell, Bannister, B\"{o}hm, Colgan, Kieslich, Loch, Mitnik, M\"{u}ller
  et~al.}}]{Schnell03}
\bibinfo{author}{\bibfnamefont{M.}~\bibnamefont{Schnell}},
  \bibinfo{author}{\bibfnamefont{G.}~\bibnamefont{Gwinner}},
  \bibinfo{author}{\bibfnamefont{N.~R.} \bibnamefont{Badnell}},
  \bibinfo{author}{\bibfnamefont{M.~E.} \bibnamefont{Bannister}},
  \bibinfo{author}{\bibfnamefont{S.}~\bibnamefont{B\"{o}hm}},
  \bibinfo{author}{\bibfnamefont{J.}~\bibnamefont{Colgan}},
  \bibinfo{author}{\bibfnamefont{S.}~\bibnamefont{Kieslich}},
  \bibinfo{author}{\bibfnamefont{S.~D.} \bibnamefont{Loch}},
  \bibinfo{author}{\bibfnamefont{D.}~\bibnamefont{Mitnik}},
  \bibinfo{author}{\bibfnamefont{A.}~\bibnamefont{M\"{u}ller}},
  \bibnamefont{et~al.}, \bibinfo{journal}{Phys. Rev. Lett.}
  \textbf{\bibinfo{volume}{91}}, \bibinfo{pages}{043001}
  (\bibinfo{year}{2003}).

\bibitem[{\citenamefont{Orban et~al.}(2010)\citenamefont{Orban, Loch, B\"{o}hm,
  and Schuch}}]{Orban10}
\bibinfo{author}{\bibfnamefont{I.}~\bibnamefont{Orban}},
  \bibinfo{author}{\bibfnamefont{S.~D.} \bibnamefont{Loch}},
  \bibinfo{author}{\bibfnamefont{S.}~\bibnamefont{B\"{o}hm}}, \bibnamefont{and}
  \bibinfo{author}{\bibfnamefont{R.}~\bibnamefont{Schuch}},
  \bibinfo{journal}{Astrophys. J.} \textbf{\bibinfo{volume}{721}},
  \bibinfo{pages}{1603} (\bibinfo{year}{2010}).

\bibitem[{\citenamefont{Schippers et~al.}(2010)\citenamefont{Schippers,
  Lestinsky, M\"uller, Savin, Schmidt, and Wolf}}]{Schippers10}
\bibinfo{author}{\bibfnamefont{S.}~\bibnamefont{Schippers}},
  \bibinfo{author}{\bibfnamefont{M.}~\bibnamefont{Lestinsky}},
  \bibinfo{author}{\bibfnamefont{A.}~\bibnamefont{M\"uller}},
  \bibinfo{author}{\bibfnamefont{D.~W.} \bibnamefont{Savin}},
  \bibinfo{author}{\bibfnamefont{E.~W.} \bibnamefont{Schmidt}},
  \bibnamefont{and} \bibinfo{author}{\bibfnamefont{A.}~\bibnamefont{Wolf}},
  \bibinfo{journal}{Int. Rev. At. Mol. Phys.} \textbf{\bibinfo{volume}{1}},
  \bibinfo{pages}{109} (\bibinfo{year}{2010}).

\bibitem[{\citenamefont{Grant et~al.}(1976)\citenamefont{Grant, Mayers, and
  Pyper}}]{Grant76}
\bibinfo{author}{\bibfnamefont{I.~P.} \bibnamefont{Grant}},
  \bibinfo{author}{\bibfnamefont{D.~F.} \bibnamefont{Mayers}},
  \bibnamefont{and} \bibinfo{author}{\bibfnamefont{N.~C.} \bibnamefont{Pyper}},
  \bibinfo{journal}{J. Phys. B} \textbf{\bibinfo{volume}{9}},
  \bibinfo{pages}{2777} (\bibinfo{year}{1976}).

\bibitem[{\citenamefont{{Crespo L\'{o}pez-Urrutia}
  et~al.}(1999)\citenamefont{{Crespo L\'{o}pez-Urrutia}, Dorn, Moshammer, and
  Ullrich}}]{Crespo99}
\bibinfo{author}{\bibfnamefont{J.~R.} \bibnamefont{{Crespo
  L\'{o}pez-Urrutia}}}, \bibinfo{author}{\bibfnamefont{A.}~\bibnamefont{Dorn}},
  \bibinfo{author}{\bibfnamefont{R.}~\bibnamefont{Moshammer}},
  \bibnamefont{and} \bibinfo{author}{\bibfnamefont{J.}~\bibnamefont{Ullrich}},
  \bibinfo{journal}{Physica Scripta} \textbf{\bibinfo{volume}{T80B}},
  \bibinfo{pages}{502} (\bibinfo{year}{1999}).

\bibitem[{\citenamefont{Epp et~al.}(2007)\citenamefont{Epp, {Crespo
  L\'{o}pez-Urrutia}, Brenner, M\"{a}ckel, Mokler, Treusch, Kuhlmann, Yurkov,
  Feldhaus, Schneider et~al.}}]{Epp07}
\bibinfo{author}{\bibfnamefont{S.~W.} \bibnamefont{Epp}},
  \bibinfo{author}{\bibfnamefont{J.~R.} \bibnamefont{{Crespo
  L\'{o}pez-Urrutia}}},
  \bibinfo{author}{\bibfnamefont{G.}~\bibnamefont{Brenner}},
  \bibinfo{author}{\bibfnamefont{V.}~\bibnamefont{M\"{a}ckel}},
  \bibinfo{author}{\bibfnamefont{P.~H.} \bibnamefont{Mokler}},
  \bibinfo{author}{\bibfnamefont{R.}~\bibnamefont{Treusch}},
  \bibinfo{author}{\bibfnamefont{M.}~\bibnamefont{Kuhlmann}},
  \bibinfo{author}{\bibfnamefont{M.~V.} \bibnamefont{Yurkov}},
  \bibinfo{author}{\bibfnamefont{J.}~\bibnamefont{Feldhaus}},
  \bibinfo{author}{\bibfnamefont{J.~R.} \bibnamefont{Schneider}},
  \bibnamefont{et~al.}, \bibinfo{journal}{Phys. Rev. Lett.}
  \textbf{\bibinfo{volume}{98}}, \bibinfo{pages}{183001}
  (\bibinfo{year}{2007}).

\bibitem[{\citenamefont{Widmann et~al.}(1995)\citenamefont{Widmann,
  Beiersdorfer, Decaux, Elliott, Knapp, Osterheld, Bitter, and
  Smith}}]{Widmann95}
\bibinfo{author}{\bibfnamefont{K.}~\bibnamefont{Widmann}},
  \bibinfo{author}{\bibfnamefont{P.}~\bibnamefont{Beiersdorfer}},
  \bibinfo{author}{\bibfnamefont{V.}~\bibnamefont{Decaux}},
  \bibinfo{author}{\bibfnamefont{S.~R.} \bibnamefont{Elliott}},
  \bibinfo{author}{\bibfnamefont{D.}~\bibnamefont{Knapp}},
  \bibinfo{author}{\bibfnamefont{A.}~\bibnamefont{Osterheld}},
  \bibinfo{author}{\bibfnamefont{M.}~\bibnamefont{Bitter}}, \bibnamefont{and}
  \bibinfo{author}{\bibfnamefont{A.}~\bibnamefont{Smith}},
  \bibinfo{journal}{Rev. Sci. Instrum.} \textbf{\bibinfo{volume}{66}},
  \bibinfo{pages}{761} (\bibinfo{year}{1995}).

\bibitem[{\citenamefont{Bitter et~al.}(1993)\citenamefont{Bitter, Hsuan, Bush,
  Cohen, Cummings, Grek, Hill, Schivell, Zarnstorff, Beiersdorfer
  et~al.}}]{Bitter93}
\bibinfo{author}{\bibfnamefont{M.}~\bibnamefont{Bitter}},
  \bibinfo{author}{\bibfnamefont{H.}~\bibnamefont{Hsuan}},
  \bibinfo{author}{\bibfnamefont{C.}~\bibnamefont{Bush}},
  \bibinfo{author}{\bibfnamefont{S.}~\bibnamefont{Cohen}},
  \bibinfo{author}{\bibfnamefont{C.~J.} \bibnamefont{Cummings}},
  \bibinfo{author}{\bibfnamefont{B.}~\bibnamefont{Grek}},
  \bibinfo{author}{\bibfnamefont{K.~W.} \bibnamefont{Hill}},
  \bibinfo{author}{\bibfnamefont{J.}~\bibnamefont{Schivell}},
  \bibinfo{author}{\bibfnamefont{M.}~\bibnamefont{Zarnstorff}},
  \bibinfo{author}{\bibfnamefont{P.}~\bibnamefont{Beiersdorfer}},
  \bibnamefont{et~al.}, \bibinfo{journal}{Phys. Rev. Lett.}
  \textbf{\bibinfo{volume}{71}}, \bibinfo{pages}{1007} (\bibinfo{year}{1993}).

\bibitem[{\citenamefont{Haan and Jacobs}(1989)}]{Haan89}
\bibinfo{author}{\bibfnamefont{S.~L.} \bibnamefont{Haan}} \bibnamefont{and}
  \bibinfo{author}{\bibfnamefont{V.~L.} \bibnamefont{Jacobs}},
  \bibinfo{journal}{Phys. Rev. A} \textbf{\bibinfo{volume}{40}},
  \bibinfo{pages}{80} (\bibinfo{year}{1989}).

\bibitem[{\citenamefont{Zimmerer et~al.}(1990)\citenamefont{Zimmerer, Gr\"{u}n,
  and Scheid}}]{Zimmerer90}
\bibinfo{author}{\bibfnamefont{P.}~\bibnamefont{Zimmerer}},
  \bibinfo{author}{\bibfnamefont{N.}~\bibnamefont{Gr\"{u}n}}, \bibnamefont{and}
  \bibinfo{author}{\bibfnamefont{W.}~\bibnamefont{Scheid}},
  \bibinfo{journal}{Phys. Lett. A} \textbf{\bibinfo{volume}{148}},
  \bibinfo{pages}{457} (\bibinfo{year}{1990}).

\bibitem[{\citenamefont{Zimmermann et~al.}(1997)\citenamefont{Zimmermann,
  Gr\"{u}n, and Scheid}}]{Zimmermann97}
\bibinfo{author}{\bibfnamefont{M.}~\bibnamefont{Zimmermann}},
  \bibinfo{author}{\bibfnamefont{N.}~\bibnamefont{Gr\"{u}n}}, \bibnamefont{and}
  \bibinfo{author}{\bibfnamefont{W.}~\bibnamefont{Scheid}},
  \bibinfo{journal}{J. Phys. B} \textbf{\bibinfo{volume}{30}},
  \bibinfo{pages}{5259} (\bibinfo{year}{1997}).

\bibitem[{\citenamefont{Eichler and Meyerhof}(1995)}]{Eichler95}
\bibinfo{author}{\bibfnamefont{J.}~\bibnamefont{Eichler}} \bibnamefont{and}
  \bibinfo{author}{\bibfnamefont{W.~E.} \bibnamefont{Meyerhof}},
  \emph{\bibinfo{title}{{Relativistic Atomic Collisions}}}
  (\bibinfo{publisher}{Academic Press}, \bibinfo{year}{1995}).

\bibitem[{\citenamefont{{Gonz\'{a}lez Mart\'inez}
  et~al.}(2006)\citenamefont{{Gonz\'{a}lez Mart\'inez}, {Crespo
  L\'{o}pez-Urrutia}, Braun, Brenner, Bruhns, Lapierre, Mironov, {Soria Orts},
  Tawara, Trinczek et~al.}}]{Gonzalez06}
\bibinfo{author}{\bibfnamefont{A.~J.} \bibnamefont{{Gonz\'{a}lez Mart\'inez}}},
  \bibinfo{author}{\bibfnamefont{J.~R.} \bibnamefont{{Crespo
  L\'{o}pez-Urrutia}}},
  \bibinfo{author}{\bibfnamefont{J.}~\bibnamefont{Braun}},
  \bibinfo{author}{\bibfnamefont{G.}~\bibnamefont{Brenner}},
  \bibinfo{author}{\bibfnamefont{H.}~\bibnamefont{Bruhns}},
  \bibinfo{author}{\bibfnamefont{A.}~\bibnamefont{Lapierre}},
  \bibinfo{author}{\bibfnamefont{V.}~\bibnamefont{Mironov}},
  \bibinfo{author}{\bibfnamefont{R.}~\bibnamefont{{Soria Orts}}},
  \bibinfo{author}{\bibfnamefont{H.}~\bibnamefont{Tawara}},
  \bibinfo{author}{\bibfnamefont{M.}~\bibnamefont{Trinczek}},
  \bibnamefont{et~al.}, \bibinfo{journal}{Phys. Rev. A}
  \textbf{\bibinfo{volume}{73}}, \bibinfo{pages}{052710}
  (\bibinfo{year}{2006}).

\bibitem[{\citenamefont{Parpia et~al.}(1996)\citenamefont{Parpia, Fischer, and
  Grant}}]{Parpia96}
\bibinfo{author}{\bibfnamefont{F.~A.} \bibnamefont{Parpia}},
  \bibinfo{author}{\bibnamefont{Fischer}}, \bibnamefont{and}
  \bibinfo{author}{\bibfnamefont{I.~P.} \bibnamefont{Grant}},
  \bibinfo{journal}{Comput. Phys. Commun.} \textbf{\bibinfo{volume}{94}},
  \bibinfo{pages}{249} (\bibinfo{year}{1996}).

\bibitem[{\citenamefont{Zakowicz et~al.}(2003)\citenamefont{Zakowicz, Harman,
  Gr\"{u}n, and Scheid}}]{Zakowicz03}
\bibinfo{author}{\bibfnamefont{S.}~\bibnamefont{Zakowicz}},
  \bibinfo{author}{\bibfnamefont{Z.}~\bibnamefont{Harman}},
  \bibinfo{author}{\bibfnamefont{N.}~\bibnamefont{Gr\"{u}n}}, \bibnamefont{and}
  \bibinfo{author}{\bibfnamefont{W.}~\bibnamefont{Scheid}},
  \bibinfo{journal}{Phys. Rev. A} \textbf{\bibinfo{volume}{68}},
  \bibinfo{pages}{042711} (\bibinfo{year}{2003}).

\bibitem[{\citenamefont{Beilmann et~al.}(2013)\citenamefont{Beilmann, Amaro,
  Bekker, Harman, {Crespo L\'opez-Urrutia}, and Tashenov}}]{Beilmann12}
\bibinfo{author}{\bibfnamefont{C.}~\bibnamefont{Beilmann}},
  \bibinfo{author}{\bibfnamefont{P.}~\bibnamefont{Amaro}},
  \bibinfo{author}{\bibfnamefont{H.}~\bibnamefont{Bekker}},
  \bibinfo{author}{\bibfnamefont{Z.}~\bibnamefont{Harman}},
  \bibinfo{author}{\bibfnamefont{J.~R.} \bibnamefont{{Crespo
  L\'opez-Urrutia}}}, \bibnamefont{and}
  \bibinfo{author}{\bibfnamefont{S.}~\bibnamefont{Tashenov}},
  \bibinfo{journal}{Phys. Scripta}  (\bibinfo{year}{2013}), \bibinfo{note}{in
  print}.

\bibitem[{\citenamefont{Knapp et~al.}(1993)\citenamefont{Knapp, Marrs,
  Schneider, Chen, Levine, and Lee}}]{Knapp93}
\bibinfo{author}{\bibfnamefont{D.~A.} \bibnamefont{Knapp}},
  \bibinfo{author}{\bibfnamefont{R.~E.} \bibnamefont{Marrs}},
  \bibinfo{author}{\bibfnamefont{M.~B.} \bibnamefont{Schneider}},
  \bibinfo{author}{\bibfnamefont{M.~H.} \bibnamefont{Chen}},
  \bibinfo{author}{\bibfnamefont{M.~A.} \bibnamefont{Levine}},
  \bibnamefont{and} \bibinfo{author}{\bibfnamefont{P.}~\bibnamefont{Lee}},
  \bibinfo{journal}{Phys. Rev. A} \textbf{\bibinfo{volume}{47}},
  \bibinfo{pages}{2039} (\bibinfo{year}{1993}).

\bibitem[{\citenamefont{Penetrante
  et~al.}(1991{\natexlab{a}})\citenamefont{Penetrante, Bardsley, Levine, Knapp,
  and Marrs}}]{Penetrante91}
\bibinfo{author}{\bibfnamefont{B.~M.} \bibnamefont{Penetrante}},
  \bibinfo{author}{\bibfnamefont{J.~N.} \bibnamefont{Bardsley}},
  \bibinfo{author}{\bibfnamefont{M.~A.} \bibnamefont{Levine}},
  \bibinfo{author}{\bibfnamefont{D.~A.} \bibnamefont{Knapp}}, \bibnamefont{and}
  \bibinfo{author}{\bibfnamefont{R.~E.} \bibnamefont{Marrs}},
  \bibinfo{journal}{Phys. Rev. A} \textbf{\bibinfo{volume}{43}},
  \bibinfo{pages}{4873} (\bibinfo{year}{1991}{\natexlab{a}}).

\bibitem[{\citenamefont{Penetrante
  et~al.}(1991{\natexlab{b}})\citenamefont{Penetrante, Bardsley, DeWitt, Clark,
  and Schneider}}]{Penetrante91sc}
\bibinfo{author}{\bibfnamefont{B.~M.} \bibnamefont{Penetrante}},
  \bibinfo{author}{\bibfnamefont{J.~N.} \bibnamefont{Bardsley}},
  \bibinfo{author}{\bibfnamefont{D.}~\bibnamefont{DeWitt}},
  \bibinfo{author}{\bibfnamefont{M.}~\bibnamefont{Clark}}, \bibnamefont{and}
  \bibinfo{author}{\bibfnamefont{D.}~\bibnamefont{Schneider}},
  \bibinfo{journal}{Phys. Rev. A} \textbf{\bibinfo{volume}{43}},
  \bibinfo{pages}{4861} (\bibinfo{year}{1991}{\natexlab{b}}).

\bibitem[{\citenamefont{Beilmann et~al.}(2010)\citenamefont{Beilmann, {Crespo
  L\'{o}pez-Urrutia}, Mokler, and Ullrich}}]{Beilmann10}
\bibinfo{author}{\bibfnamefont{C.}~\bibnamefont{Beilmann}},
  \bibinfo{author}{\bibfnamefont{J.~R.} \bibnamefont{{Crespo
  L\'{o}pez-Urrutia}}}, \bibinfo{author}{\bibfnamefont{P.~H.}
  \bibnamefont{Mokler}}, \bibnamefont{and}
  \bibinfo{author}{\bibfnamefont{J.}~\bibnamefont{Ullrich}},
  \bibinfo{journal}{J. Instrum.} \textbf{\bibinfo{volume}{5}},
  \bibinfo{pages}{C09002} (\bibinfo{year}{2010}).

\bibitem[{\citenamefont{Mokler and Folkmann}(1978)}]{Mokler78}
\bibinfo{author}{\bibfnamefont{P.~H.} \bibnamefont{Mokler}} \bibnamefont{and}
  \bibinfo{author}{\bibfnamefont{F.}~\bibnamefont{Folkmann}},
  \emph{\bibinfo{title}{{Structure and Collisions of Ions and Atoms}}},
  vol.~\bibinfo{volume}{5} of \emph{\bibinfo{series}{Topics in Current
  Physics}} (\bibinfo{publisher}{Springer-Verlag}, \bibinfo{address}{Berlin},
  \bibinfo{year}{1978}).

\bibitem[{\citenamefont{Zimmerer et~al.}(1991)\citenamefont{Zimmerer, Gr\"un,
  and Scheid}}]{Zimmerer91}
\bibinfo{author}{\bibfnamefont{P.}~\bibnamefont{Zimmerer}},
  \bibinfo{author}{\bibfnamefont{N.}~\bibnamefont{Gr\"un}}, \bibnamefont{and}
  \bibinfo{author}{\bibfnamefont{W.}~\bibnamefont{Scheid}},
  \bibinfo{journal}{J. Phys. B} \textbf{\bibinfo{volume}{24}},
  \bibinfo{pages}{2633} (\bibinfo{year}{1991}).

\bibitem[{\citenamefont{Watanabe et~al.}(2002)\citenamefont{Watanabe, Currell,
  Kuramoto, Ohtani, O'Rourke, and Tong}}]{Watanabe02}
\bibinfo{author}{\bibfnamefont{H.}~\bibnamefont{Watanabe}},
  \bibinfo{author}{\bibfnamefont{F.~J.} \bibnamefont{Currell}},
  \bibinfo{author}{\bibfnamefont{H.}~\bibnamefont{Kuramoto}},
  \bibinfo{author}{\bibfnamefont{S.}~\bibnamefont{Ohtani}},
  \bibinfo{author}{\bibfnamefont{B.~E.} \bibnamefont{O'Rourke}},
  \bibnamefont{and} \bibinfo{author}{\bibfnamefont{X.~M.} \bibnamefont{Tong}},
  \bibinfo{journal}{J. Phys. B} \textbf{\bibinfo{volume}{35}},
  \bibinfo{pages}{5095} (\bibinfo{year}{2002}).

\bibitem[{\citenamefont{Kavanagh et~al.}(2010)\citenamefont{Kavanagh, Watanabe,
  Li, O'Rourke, Tobiyama, Nakamura, McMahon, Yamada, Ohtani, and
  Currell}}]{Kavanagh10}
\bibinfo{author}{\bibfnamefont{A.~P.} \bibnamefont{Kavanagh}},
  \bibinfo{author}{\bibfnamefont{H.}~\bibnamefont{Watanabe}},
  \bibinfo{author}{\bibfnamefont{Y.~M.} \bibnamefont{Li}},
  \bibinfo{author}{\bibfnamefont{B.~E.} \bibnamefont{O'Rourke}},
  \bibinfo{author}{\bibfnamefont{H.}~\bibnamefont{Tobiyama}},
  \bibinfo{author}{\bibfnamefont{N.}~\bibnamefont{Nakamura}},
  \bibinfo{author}{\bibfnamefont{S.}~\bibnamefont{McMahon}},
  \bibinfo{author}{\bibfnamefont{C.}~\bibnamefont{Yamada}},
  \bibinfo{author}{\bibfnamefont{S.}~\bibnamefont{Ohtani}}, \bibnamefont{and}
  \bibinfo{author}{\bibfnamefont{F.~J.} \bibnamefont{Currell}},
  \bibinfo{journal}{Phys. Rev. A} \textbf{\bibinfo{volume}{81}},
  \bibinfo{pages}{022712} (\bibinfo{year}{2010}).

\bibitem[{\citenamefont{Simon et~al.}(2010)\citenamefont{Simon, {Crespo
  L\'{o}pez-Urrutia}, Beilmann, Schwarz, Harman, Epp, Schmitt, Baumann, Behar,
  Bernitt et~al.}}]{Simon10a}
\bibinfo{author}{\bibfnamefont{M.~C.} \bibnamefont{Simon}},
  \bibinfo{author}{\bibfnamefont{J.~R.} \bibnamefont{{Crespo
  L\'{o}pez-Urrutia}}},
  \bibinfo{author}{\bibfnamefont{C.}~\bibnamefont{Beilmann}},
  \bibinfo{author}{\bibfnamefont{M.}~\bibnamefont{Schwarz}},
  \bibinfo{author}{\bibfnamefont{Z.}~\bibnamefont{Harman}},
  \bibinfo{author}{\bibfnamefont{S.~W.} \bibnamefont{Epp}},
  \bibinfo{author}{\bibfnamefont{B.~L.} \bibnamefont{Schmitt}},
  \bibinfo{author}{\bibfnamefont{T.~M.} \bibnamefont{Baumann}},
  \bibinfo{author}{\bibfnamefont{E.}~\bibnamefont{Behar}},
  \bibinfo{author}{\bibfnamefont{S.}~\bibnamefont{Bernitt}},
  \bibnamefont{et~al.}, \bibinfo{journal}{Phys. Rev. Lett.}
  \textbf{\bibinfo{volume}{105}}, \bibinfo{pages}{183001}
  (\bibinfo{year}{2010}).

\bibitem[{\citenamefont{Aksela et~al.}(1996)\citenamefont{Aksela, Aksela, and
  Kabachnik}}]{Aksela96a}
\bibinfo{author}{\bibfnamefont{H.}~\bibnamefont{Aksela}},
  \bibinfo{author}{\bibfnamefont{A.}~\bibnamefont{Aksela}}, \bibnamefont{and}
  \bibinfo{author}{\bibfnamefont{N.}~\bibnamefont{Kabachnik}},
  \emph{\bibinfo{title}{{Resonant and Nonresonant Auger Recombination}}}
  (\bibinfo{publisher}{Plenum}, \bibinfo{address}{New York},
  \bibinfo{year}{1996}), p. \bibinfo{pages}{401}, Physics of Atoms and
  Molecules.

\bibitem[{\citenamefont{Uiberacker et~al.}(2007)\citenamefont{Uiberacker,
  Uphues, Schultze, Verhoef, Yakovlev, Kling, Rauschenberger, Kabachnik,
  Schroder, Lezius et~al.}}]{Uiberacker07}
\bibinfo{author}{\bibfnamefont{M.}~\bibnamefont{Uiberacker}},
  \bibinfo{author}{\bibfnamefont{T.}~\bibnamefont{Uphues}},
  \bibinfo{author}{\bibfnamefont{M.}~\bibnamefont{Schultze}},
  \bibinfo{author}{\bibfnamefont{A.~J.} \bibnamefont{Verhoef}},
  \bibinfo{author}{\bibfnamefont{V.}~\bibnamefont{Yakovlev}},
  \bibinfo{author}{\bibfnamefont{M.~F.} \bibnamefont{Kling}},
  \bibinfo{author}{\bibfnamefont{J.}~\bibnamefont{Rauschenberger}},
  \bibinfo{author}{\bibfnamefont{N.~M.} \bibnamefont{Kabachnik}},
  \bibinfo{author}{\bibfnamefont{H.}~\bibnamefont{Schroder}},
  \bibinfo{author}{\bibfnamefont{M.}~\bibnamefont{Lezius}},
  \bibnamefont{et~al.}, \bibinfo{journal}{Nature}
  \textbf{\bibinfo{volume}{446}}, \bibinfo{pages}{627} (\bibinfo{year}{2007}).

\bibitem[{\citenamefont{Fritzsche et~al.}(2009)\citenamefont{Fritzsche,
  Surzhykov, and St\"ohlker}}]{Fritzsche2009}
\bibinfo{author}{\bibfnamefont{S.}~\bibnamefont{Fritzsche}},
  \bibinfo{author}{\bibfnamefont{A.}~\bibnamefont{Surzhykov}},
  \bibnamefont{and}
  \bibinfo{author}{\bibfnamefont{T.}~\bibnamefont{St\"ohlker}},
  \bibinfo{journal}{Phys. Rev. Lett.} \textbf{\bibinfo{volume}{103}},
  \bibinfo{pages}{113001} (\bibinfo{year}{2009}).

\bibitem[{\citenamefont{M\"{u}ller et~al.}(2010)\citenamefont{M\"{u}ller,
  Voitkiv, {Crespo L\'{o}pez-Urrutia}, and Harman}}]{MuellerC10}
\bibinfo{author}{\bibfnamefont{C.}~\bibnamefont{M\"{u}ller}},
  \bibinfo{author}{\bibfnamefont{A.~B.} \bibnamefont{Voitkiv}},
  \bibinfo{author}{\bibfnamefont{J.~R.} \bibnamefont{{Crespo
  L\'{o}pez-Urrutia}}}, \bibnamefont{and}
  \bibinfo{author}{\bibfnamefont{Z.}~\bibnamefont{Harman}},
  \bibinfo{journal}{Phys. Rev. Lett.} \textbf{\bibinfo{volume}{104}},
  \bibinfo{pages}{233202} (\bibinfo{year}{2010}).

\bibitem[{\citenamefont{Voitkiv and Najjari}(2010)}]{Voitkiv10}
\bibinfo{author}{\bibfnamefont{A.~B.} \bibnamefont{Voitkiv}} \bibnamefont{and}
  \bibinfo{author}{\bibfnamefont{B.}~\bibnamefont{Najjari}},
  \bibinfo{journal}{Phys. Rev. A} \textbf{\bibinfo{volume}{82}},
  \bibinfo{pages}{052708} (\bibinfo{year}{2010}).

\end{thebibliography}
\end{document}